\documentclass[aps,pre,twocolumn,showpacs,floatfix,groupedaddress]{revtex4}

\usepackage{latexsym}
\usepackage{amsmath}
\usepackage{amsthm}
\usepackage{amssymb}
\usepackage{amscd}
\usepackage{bm}

\usepackage{graphicx}
\usepackage{epsfig}
\usepackage{epstopdf}
\usepackage[caption=false]{subfig}
\usepackage{placeins}

\newtheorem{theorem}{Theorem}
\newtheorem{conj}[theorem]{Conjecture}

\renewcommand{\vec}[1]{\boldsymbol{#1}}

\providecommand{\nvec}[1]{\hat{\boldsymbol{#1}}}

\providecommand{\diff}{\mathrm{d}}
\providecommand{\deriv}[3][ ]{\frac{\diff^{#1}#2}{\diff #3^{#1}}}

\providecommand{\norm}[1]{\left\Vert#1\right\Vert}

\providecommand{\paren}[1]{( #1 )}

\providecommand{\eqref}[1]{(\ref{#1})}

\begin{document}

\title{Topological Similarity of Random Cell Complexes and Applications}

\author{B. Schweinhart}
\email{bschwein@math.princeton.edu}
\affiliation{Center of Mathematical Sciences and Applications, Harvard University, Cambridge, Massachusetts 02138, USA.}
\author{J. K. Mason}
\email{jeremy.mason@boun.edu.tr}
\affiliation{Bo\u{g}azi\c{c}i University, Bebek, Istanbul 34342, TR.}
\author{R. D. MacPherson}
\email{rdm@math.ias.edu}
\affiliation{School of Mathematics, Institute for Advanced Study, Princeton, New Jersey 08540, USA.}

\begin{abstract}
Although random cell complexes occur throughout the physical sciences, there does not appear to be a standard way to quantify their statistical similarities and differences. The various proposals in the literature are usually motivated by the analysis of particular physical systems and do not necessarily apply to general situations. The central concepts in this paper---the swatch and the cloth---provide a description of the local topology of a cell complex that is general (any physical system that can be represented as a cell complex is admissible) and complete (any statistical question about the local topology can be answered from the cloth). Furthermore, this approach allows a distance to be defined that measures the similarity of the local topology of two cell complexes. The distance is used to identify a steady state of a model grain boundary network, to quantify the approach to this steady state, and to show that the steady state is independent of the initial conditions. The same distance is then employed to show that the long-term properties in simulations of a specific model of a dislocation network does not depend on the implementation of dislocation intersections.
\end{abstract}


\maketitle

\section{Introduction}
\label{sec_introduction}

Random cell complexes (defined below) abound at all length scales in the physical sciences. Specifically with regard to materials science, examples include the contact graph of atoms in a metallic glass \cite{2006sheng} and the covalent bonds in an oxide glass \cite{1932zachariasen} at the atomic scale, dense dislocation networks \cite{2009motz} and the bonding of cross-linked polymers \cite{2013smallenburg} at the nanometer scale, and the grain boundary network in a polycrystal \cite{2010rowenhorst} and the cells in a metallic foam \cite{2001banhart} at the micrometer scale. One feature of all of these systems is that they defy characterization by the usual approach used in crystallography, that is, by the identification of a periodic unit and the classification of defects as deviations from periodicity. Nevertheless, some means of characterization is clearly necessary for these systems to be classified and eventually engineered.

Using metallic glass as a specific example, the arrangement of atoms appears to be homogenous and isotropic on average, and from this standpoint is quite simple. Yet, some recent experimental results \cite{2007swallen} indicate that differences in the preparation of samples with the same composition can result in measurably different mechanical properties. This suggests the presence of small variations in the atomic arrangements \cite{2013singh}, though at present the nature of these variations and the means to measure them remain unclear. This paper contends that the situation is similar to that of the characterization of crystalline materials before the advent of crystallography; the analysis of these systems would be vastly simplified by introducing language and concepts detailed enough to provide an accurate description, and yet abstract enough to apply to many different situations.

Numerous attempts have been made to introduce such a language already. The granocentric model \cite{2009clusel,2010corwin} numerically predicts the distributions of local quantities around a sphere in a polydisperse sphere packing. Shell distance is the minimum number of faces that must be crossed to go from the interior of one cell to the interior of a second cell in a cell structure \cite{1996szeto,1996asteA}. Ring statistics consider the lengths of the shortest closed paths through the network of bonds in a disordered covalent glass \cite{1993rino,2010leroux}. Homology theory is a related but more general approach that characterizes holes of arbitrary dimension \cite{2012macpherson,2013kramar,2014nakamura}. The Randi\`{c} index measures the degree of branching in organic molecules by considering the types of edges in the adjacency graph of the atoms \cite{1975randic}. Percolation theory is concerned with connected clusters of occupied vertices or edges in a graph, and particularly with the appearance of a unique infinite cluster \cite{1994stauffer,1996jacobs}. A hyperuniform distribution of points has the property that infinite-wavelength density fluctuations are absent \cite{2003torquato,2011zachary}.

While certainly not exhaustive, this selection from the available literature is intended to show that existing approaches generally have two limitations. First, they are often motivated by and formulated for a specific situation, and cannot be applied generally. Second, to the extent of our knowledge, none of them offers a complete characterization of the local topology of a physical system. That is, they do not allow the local structure to be reconstructed exactly up to a geometric deformation.

A cell complex is general in the sense that all of the physical systems described above (and many others) can be represented by means of one, and hence is used as a common point of departure in the following.  A cell complex is composed of cells, where a $0$-cell corresponds to a point, a $1$-cell to a line segment, a $2$-cell to a disk, and a $3$-cell to a ball. A cell complex is constructed by placing the necessary $0$-cells, attaching the endpoints of deformed $1$-cells to the $0$-cells, attaching the boundaries of deformed $2$-cells to the $1$-cells, and attaching the surfaces of deformed $3$-cells to the $2$-cells. The cells are attached without any twisting or tearing. This allows the description of, e.g., fused quartz with the silicon atoms as $0$-cells and the oxygen atoms as $1$-cells, soap foams with junctions as $0$-cells, edges as $1$-cells, and surfaces as $2$-cells, and polycrystals with nodes as $0$-cells, triple lines as $1$-cells, boundaries as $2$-cells, and grains as $3$-cells \footnote{In the mathematical literature, the cell complexes defined here are referred to as regular cell complexes.}.

The purpose of this paper is to suggest that the swatch and the cloth serve as useful alternatives for the description of a cell complex \cite{2012masonB}. A swatch completely characterizes the local topology of a small region of the cell complex, and a cloth indicates the frequencies of the various swatch types occurring in the cell complex. The advantages of this technique are that the description is complete (any question about the local topology of the cell complex can be answered from the cloth), that the description is general (any physical system that can be represented as a cell complex is admissible), and that the description allows the construction of a distance that quantifies the similarity of the statistical topology of two different cell complexes. This distance is useful, e.g., when considering the convergence of numerical studies, when comparing simulated cell complexes with experimental ones, when quantifying the variability of cell complexes generated in a particular way, or when iteratively modifying some cell complex to reach an intended target. For instance, this paper uses the distance to define a notion of convergence to a topological steady state, and to quantify the approach of our numerical simulations to this steady state. This should be contrasted with the usual approach of following several arbitrarily chosen quantities to identify convergence.

\begin{figure}
\center
\subfloat[]{%
	\label{triangle_complex}{%
		\includegraphics[height=2.2cm]{%
			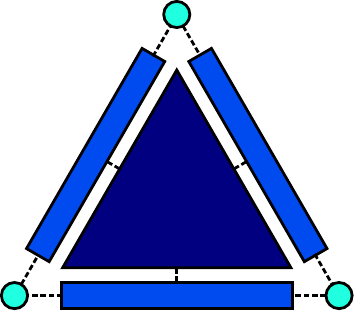}}}
\hspace{12pt}
\subfloat[]{%
	\label{triangle_graph}{%
		\includegraphics[height=2cm]{%
			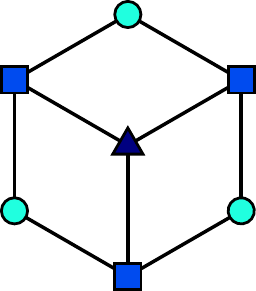}}}
\caption{\label{triangle}(a) Representation of a triangle as a cell complex by three $0$-cells, three $1$-cells, and one $2$-cell. (b) Representation of a triangle as a graph with seven vertices, nine edges, and three vertex types.}
\end{figure}

The description of a physical system by a cell complex is further refined by transforming the cell complex into an equivalent adjacency graph, where a graph is composed of a set of vertices connected by edges. Every vertex of the graph corresponds to a cell of the cell complex, and every edge in the graph corresponds to two incident cells in the cell complex whose dimensions differ by one. For the purpose of illustration, the cell complex of a triangle in Figure \ref{triangle_complex} is equivalent to the adjacency graph in Figure \ref{triangle_graph}. Notice that this requires the introduction of three vertex types; circle, square and triangle vertices in the graph correspond to cells of dimension $0$, $1$, and $2$ in the cell complex. We will follow this convention throughout the paper. More generally, the representation of a cell complex by a graph allows the mathematical machinery of graph theory to be used, and the description of swatches in Section \ref{sec_swatches}, of cloths in Section \ref{sec_cloths}, and of cell complex similarity in Section \ref{sec_metric} to be applied with relatively few modifications to graphs that appear in a variety of subjects. 

The utility of this approach is illustrated by means of simulations of two different physical systems. The first is a two-dimensional grain boundary network that develops by a process of normal grain growth, as discussed in Section \ref{sec_grain_growth}. This system is modeled by a set of grain boundary edges with three boundary edges meeting at every triple junction, and the same mobility and energy per unit length for every boundary edge. The Turnbull relation \cite{1951turnbull} governs the motion of the boundary edges, and is equivalent to evolution by curvature flow. The explicit form of the equations of motion depends on the properties of the triple junctions. Different formulations are derived for the cases of finite and infinite triple junction mobilities. Simulation results in Section \ref{sec_steady_state} provide evidence of a statistical steady state where all statistical quantities relating to the local topology of the grain boundary network converge, which is independent of the initial conditions and is the same for the two sets of equations. The distance on cell complexes introduced in Section \ref{sec_metric} is instrumental in making meaningful comparison of the grain boundary networks possible.

The second simulated system is a dislocation network \cite{1982hirth} in a material with no surface tractions during the process of recovery. This system is modeled as a set of dislocation edges with three edges meeting at every edge endpoint, and with a constant energy per dislocation line length; that is, all dislocation interactions are neglected and only the self-energy of the dislocations is retained. Evolution occurs by energy minimization with the same kinetics for dislocation glide and climb, resulting in curvature flow and a reduction in total line length. Our belief is that this severe simplification is justified by the need to simulate networks containing millions of edges (where the calculation of the long-range stress fields would be prohibitively expensive) to reduce the statistical error, and by our emphasis on the network topology rather than on the physics of a specific deformation process. This system is also mathematically interesting independent of the physical interpretation, and is discussed in more detail in Section \ref{sec_dislocations}. As with the grain boundary network, there appears to be a statistical steady state where all statistical quantities relating to the local topology of the dislocation network converge, and where the steady state is independent of the initial conditions. Remarkably, the long-term topology of the dislocation network seems to be invariant as to whether dislocations intersect and leave behind an adjoining edge or merely pass through one another. This suggests that in situations where only the steady state configuration is desired, substantial computational savings can be realized without loss of accuracy by not implementing one of the allowed topological events.

\section{Swatches and Local Topology}
\label{sec_swatches}

Given a cell complex, a swatch is the portion of the cell complex in a region around some specified vertex. This is motivated by the observation that random cell complexes are often compared using the frequencies of local configurations of cells. A swatch provides a definition of local configurations that completely describes the local topology and is agnostic to the details of the physical system, and therefore is a suitable basis for the current approach.

The definition of a swatch begins with the selection of a central vertex known as the root. For swatches to be directly comparable, all roots should have the same vertex type; that is, all swatches should be centered on cells of the same dimension in the underlying cell complex. The convention followed throughout this paper is for the roots to be on vertices of the graph corresponding to 0-cells of the cell complex (indicated by circles in the figures), though this is not a general requirement.

After selection of a root, an integer known as the radius of the swatch is chosen to specify the extent of the local configuration. The radius is measured using the canonical graph distance; a swatch of radius $r$ includes all of the vertices that can be reached from the root by traversing no more than $r$ edges of the adjacency graph. Notice that this allows the swatch to contain as much information about the local configuration as is necessary to measure a given property of interest, provided that the property depends only on the local topology.

\begin{figure}
\center
\subfloat[]{%
	\label{free_swatch}{%
		\includegraphics[height=4cm]{%
			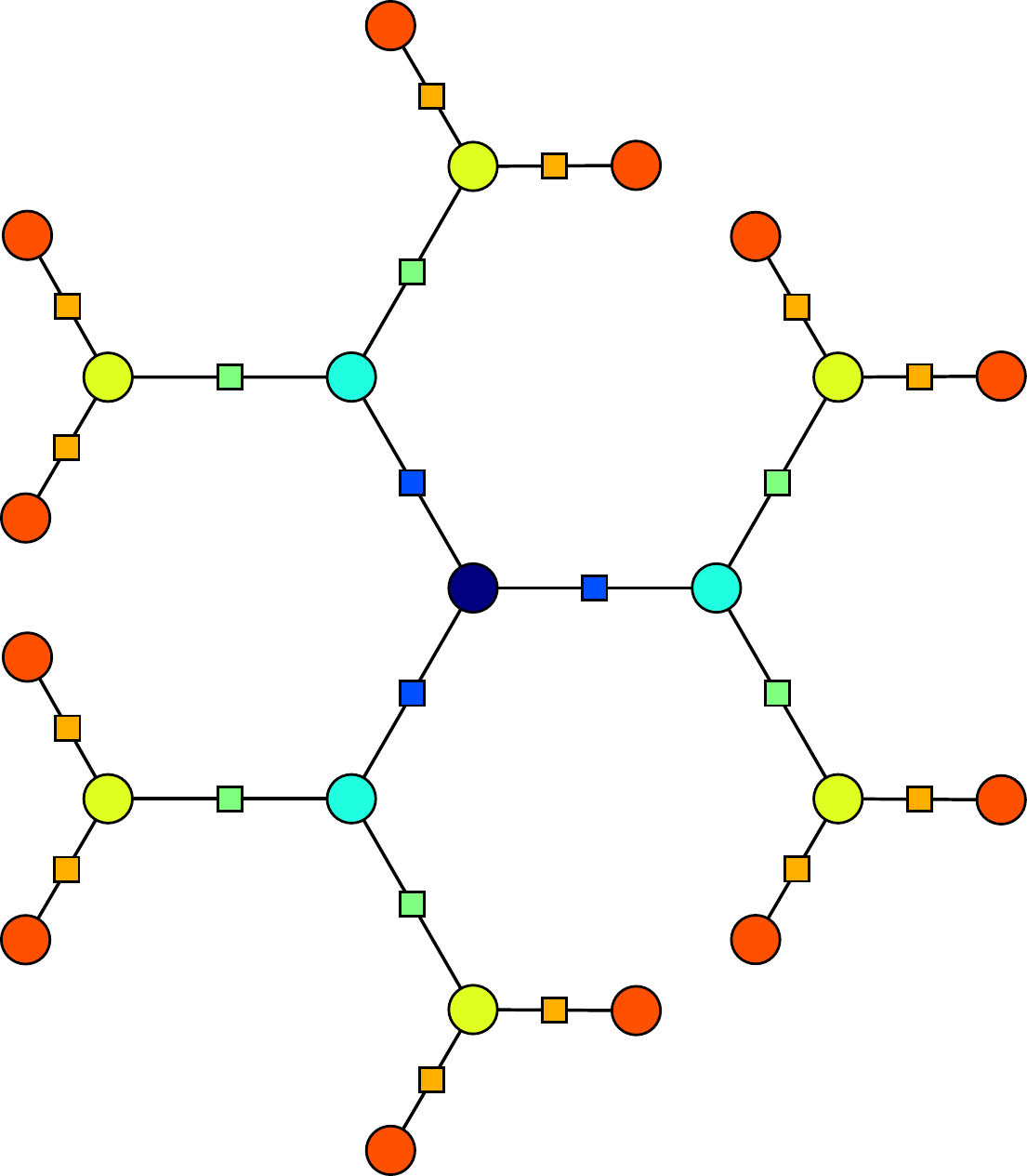}}}
\hspace{12pt}
\subfloat[]{%
	\label{not_free_swatch}{%
		\includegraphics[height=3.83cm]{%
			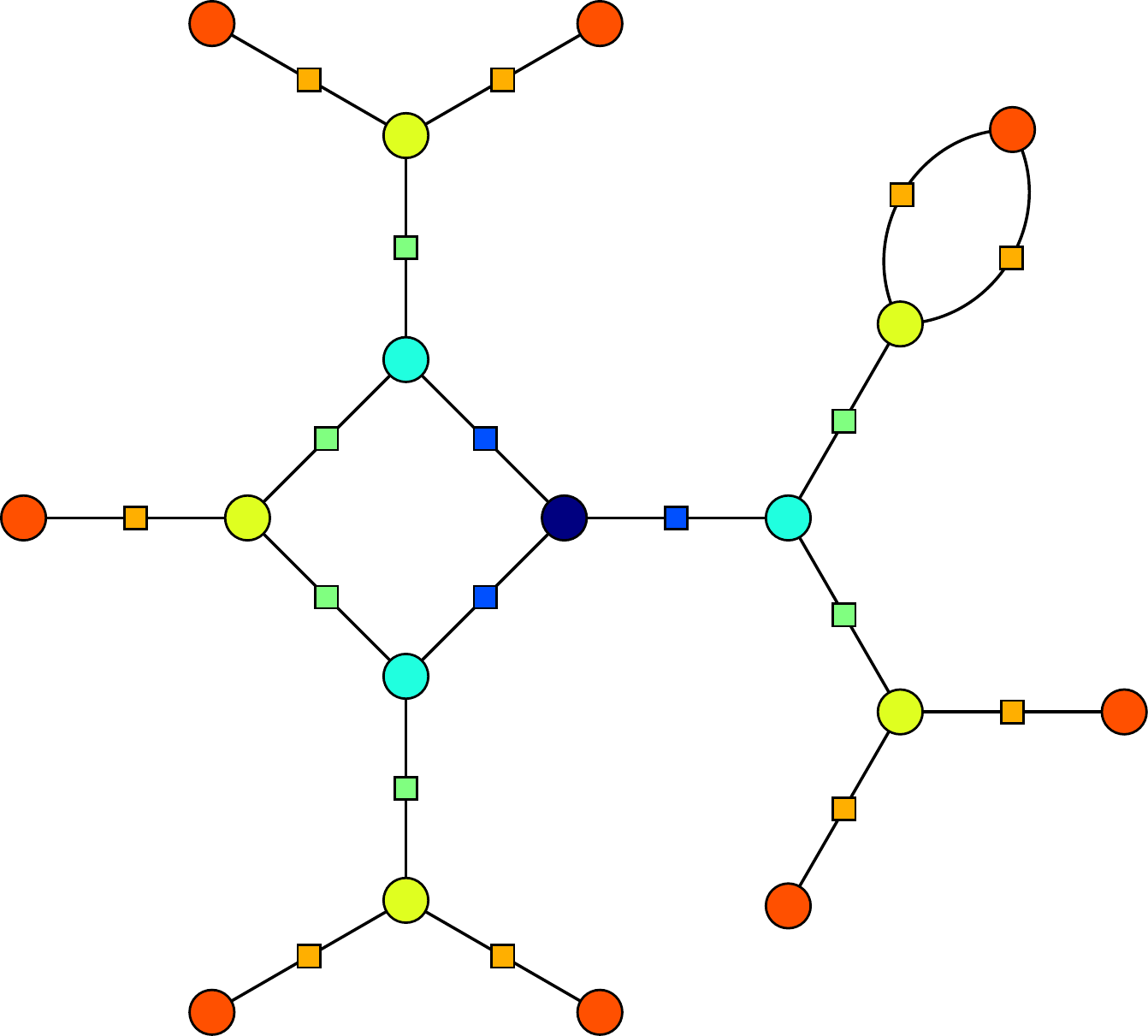}}}
\caption{\label{swatch_examples}Swatches of radius six in a cell complex containing only $0$- and $1$-cells. The vertex color indicates distance from the root, with the root colored dark blue. (a) A free swatch where there is a single path from the root to any given vertex. (b) A swatch that contains a cycle of length four and a cycle of length two.}
\end{figure}

Figure \ref{swatch_examples} provides several examples of swatches constructed from one of the grain boundary networks described in Section \ref{sec_introduction}. The root is colored dark blue, and the color of the surrounding vertices indicates distance from the root. Grain boundaries correspond to square vertices in the graph, and triple junctions correspond to circle vertices in the graph. Physical constraints require square vertices to be of degree two and circle vertices to be of degree three, where the degree of a vertex is the number of connecting edges. The most significant difference between the swatches in Figures \ref{free_swatch} and \ref{not_free_swatch} is related to the presence of cycles, or to closed paths along the edges of the graph. The free swatch in Figure \ref{free_swatch} is distinguished by the absence of cycles, while the swatch in Figure \ref{not_free_swatch} has one cycle of length four and one cycle of length two.

A distinct advantage of defining a swatch using the language of graph theory is the compact computational representation that this affords. Let the vertices of the swatch be labeled by consecutive integers from $1$ to $n$. The vertex types can be stored in an integer array of length $n$, and the edges of the swatch can be stored in an $n \times n$ adjacency matrix where the entry in the $i$th row and $j$th column is $1$ if the $i$th and $j$th vertices share an edge, and is $0$ otherwise. While this is already sufficient to reconstruct the swatch, there remains an issue of uniqueness; rearranging the vertex labels usually results in a different (but equivalent) adjacency matrix for the swatch. This complicates the comparison of two swatches with different roots since the adjacency matrices of the swatches cannot be directly compared without simultaneously considering all permutations of the vertex labels.

This difficulty is resolved by finding a canonical labeling for every swatch. When two canonically-labeled swatches have the same vertex types and adjacency matrices, the swatches describe regions with the same local topology and are said to belong to the same swatch type. Conversely, two canonically-labeled swatches with different vertex types or adjacency matrices must describe regions with different local topologies, and hence belong to different swatch types. This reduces the classification of swatches by swatch type to the problem of finding a canonical labeling for a graph. While an algorithm that performs well in all cases is still not known, the program {\it nauty} is able to find canonical labellings quite efficiently in practice \cite{2014mckay}.

\section{Cloths and Statistical Topology}
\label{sec_cloths}

A swatch provides a complete description of the local topology around a specific root of an adjacency graph, but not of the statistical local topology of the cell complex as a whole. This suggests that a swatch of radius $r$ be constructed around every root, resulting in an ensemble of swatches. The probability that a randomly selected swatch of radius $r$ belongs to a given swatch type is known as the swatch frequency, and the set of all swatch frequencies for all values of the radius is known as the cloth. The cloth characterizes the local topology of the cell complex in the sense of determining the probability of appearance of any local configuration, as well as of prescribing all local topological properties of the cell complex (as defined toward the end of Section \ref{sec_metric}).

The idea of the cloth is related to that of local isomorphism of quasicrystals \cite{1986levine}. Two quasicrystals are considered to be locally isomorphic when they can be made to coincide exactly over an arbitrarily large region by a relative translation, or equivalently, when every atomic arrangement in one occurs somewhere in the other. This is physically significant since two quasicrystals that are locally isomorphic have the same diffraction pattern. However, this description does not allow a meaningful analysis when two quasicrystals are not locally isomorphic. By contrast, a cloth provides the information about the frequencies of every swatch type of every radius. This allows not only the occurrence but the relative rates of local bond topologies to be compared, and the extent to which the quasicrystals fails to be locally isomorphic to be measured.

The cloth is composed of levels, each containing swatch frequencies for the corresponding value of the radius. Notice that level $r$ of the cloth contains strictly more information than all levels ${s < r}$ of the cloth. This follows from the observation that a swatch of radius $r$ can be restricted to a swatch of radius $s$ with the same root (known as a subswatch) by excluding all vertices further than distance $s$ from the root. The swatch types comprising level $r$ of the cloth therefore contain strictly more information than the swatch types comprising level $s$ of the cloth, and the description of the cell complex becomes progressively more complete as the level increases.

\begin{figure}
\includegraphics[width=0.95\columnwidth]{%
	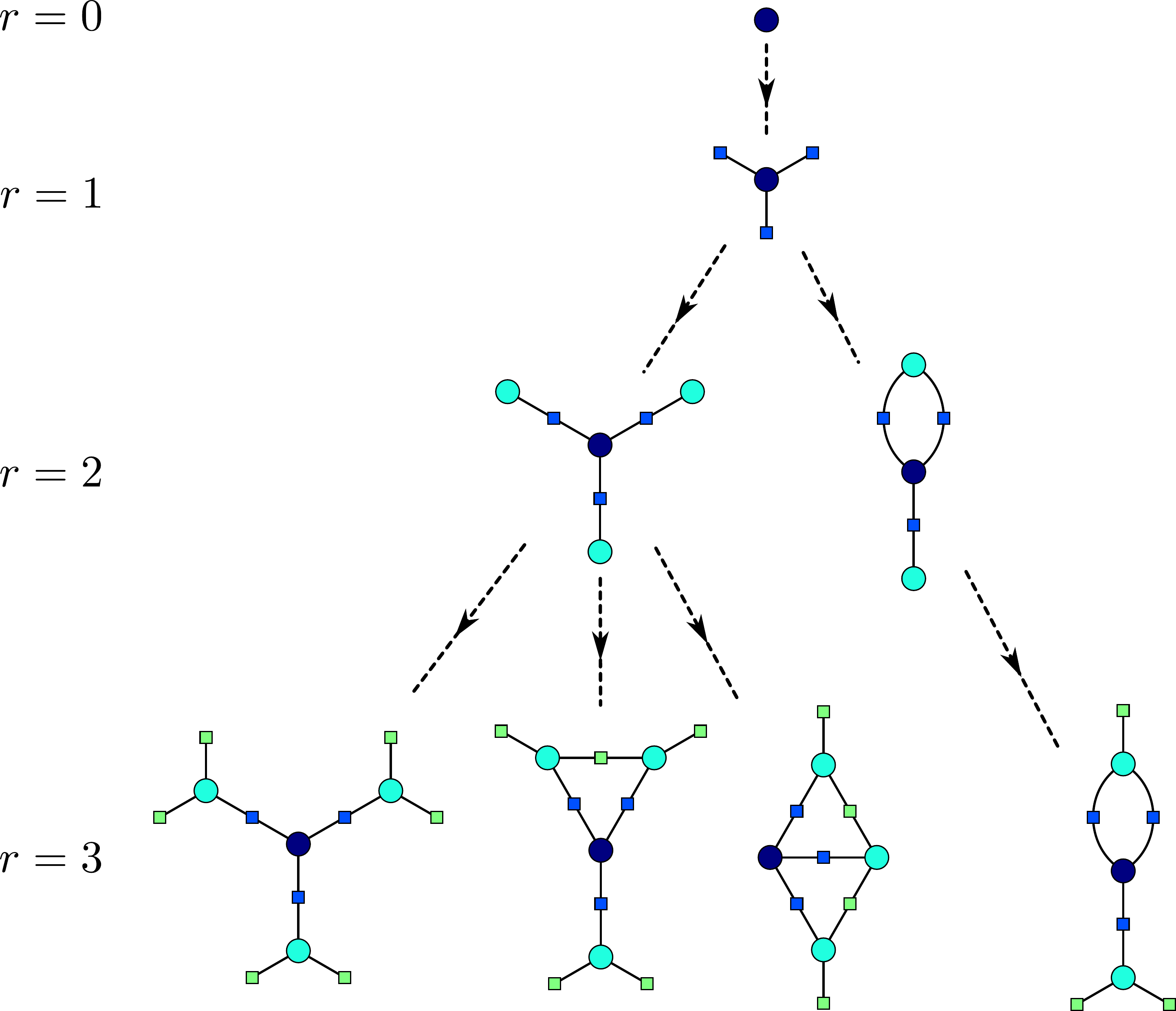}
\caption{\label{tree_of_swatches}Levels 0 to 3 of the tree of swatch types, subject to the conditions that circle vertices be of degree three, square vertices be of degree two, and the graph be 2-connected.}
\end{figure}

The relationship between different levels of the cloth can be more clearly expressed by means of a tree of swatch types, as in Figure \ref{tree_of_swatches}. Level $r$ of the tree is composed of all swatch types of radius $r$ that are compatible with the constraints of the physical system. Whenever a swatch type of radius ${r - 1}$ is a subswatch of a swatch type of radius $r$, they are connected by an edge. Since there may be more than one swatch type of radius $r$ that satisfies this condition, the tree repeatedly branches with increasing level. Figure \ref{tree_of_swatches} is specialized to the case of the grain boundary network described in Section \ref{sec_introduction} where, apart from the restrictions on the degrees of the vertex types, the adjacency graph satisfies the further condition of being 2-connected \footnote{A graph is $2$-connected when the removal of any vertex does not disconnect the graph.}.

The information in a cloth is equivalent to an assignment of swatch frequencies to each of the swatch types of the tree, subject to the condition that the frequencies on any level sum to one. This description helps to clarify the relationship of swatch frequencies on distinct levels; supposing that $S$ is a swatch type, the frequency of $S$ is equal to the sum of the frequencies of all the descendants of $S$ on any subsequent level. Since a swatch frequency is directly proportional to the number of root vertices around which swatches of the given type occur, this is a direct consequence of $S$ being a subswatch of all of the descendants of $S$.

There are several practical concerns that arise when measuring the swatch frequencies of a finite cell complex. First, the measured swatch frequencies will converge with increasing system size only when the cell complex is statistically homogeneous. That is, any statistical feature measured within a bounded region converges to a definite limit as the region's volume increases, and the limiting value is independent of the region's position within the cell complex \cite{2012masonB}. Second, experience shows that the number of swatch types grows exceedingly quickly as a function of the level number. There may be only a single occurrence of many swatch types for radii where there are more swatch types than roots, introducing substantial sampling errors. Our efforts to reduce these errors led to the adoption of the simplified dislocation network model discussed in Section \ref{sec_dislocations} as a means to increase simulation size, given the available computational resources.

\section{Similarity and Convergence of Cell Complexes}
\label{sec_metric}

Consider a sequence of cell complexes of increasing size whose cloths become ever more similar. For example, one could generate a sequence of steady state configurations of some dynamical process. It is intuitive to think of these cell complexes as approaching an infinite, universal state. This section makes that notion rigorous by introducing a distance on cell complexes and constructing limit objects corresponding to convergent cell complex sequences. The existence of these limit objects provides strong theoretical backing for the experimental results discussed in Sections \ref{sec_steady_state}. The distance can be used for many purposes, including the comparison of cell complexes arising from different processes. Note that the mathematical results in this section require that the cell complexes in the sequence have adjacency graphs with uniformly bounded degree. 

The distance on cell complexes will make use of a preliminary distance on swatches. Let the largest common subswatch of two swatches be the swatch of largest radius that is a subswatch of both. The distance between two swatches is defined as the reciprocal of the number of vertices in the largest common subswatch, or zero if the swatches are the same. For example, the largest radius for which the swatch in Figure \ref{not_free_swatch} is free is $r = 3$, and the distance to the free swatch in Figure \ref{free_swatch} is $1/13$.

Having introduced a distance on swatches, the earth mover's distance is used to define a family of distances on cell complexes. The earth mover's distance is equal to the minimum cost of transforming one probability distribution on swatch types into a different probability distribution on the same swatch types. The transformation is performed by transferring probability mass between swatch types, with the overall cost given by the sum of the costs of the individual operations. The cost of an operation is the probability mass transferred times the distance between the two swatch types \cite{1781Monge,1998Rubner}. Given two cell complexes $C_1$ and $C_2$, let $d_r\paren{C_1, C_2}$ equal the earth mover's distance between probability distributions on swatches of radius $r$ induced by the two cell complexes. Note that $d_r$ is uniformly bounded and non-decreasing in $r$, and that it stabilizes for some finite $r$ if the cell complexes are finite. The limit distance on cell complexes is defined as the limit of $d_r$ with increasing $r$, or
\begin{equation}
d\paren{C_1,C_2}=\lim_{r\rightarrow\infty}d_r\paren{C_1,C_2}. \nonumber
\end{equation}

Let $C_1, C_2, \ldots = \left\{ C_i \right\}$ be a sequence of cell complexes that it is a Cauchy sequence in the distance $d$. That is, the elements of the sequence all become arbitrarily close above some sufficiently large $i$. This condition is equivalent to the convergence of all swatch frequencies, and implies the convergence of other important properties as well.  A key mathematical result of Benjamini and Schramm \cite{2001benjamini,2012lovasz} is that the sequence $\left\{C_i\right\}$ may be associated with a universal limit object $\sigma$. The limit object is not a cell complex itself, but is instead a probability distribution on the space $\mathcal{C}^\bullet$ of countably infinite, connected cell complexes with a specified root (a root must be specified because swatches are inherently rooted). Sampling from $\sigma$ may be viewed as sampling a random configuration from the universal state that the cell complex sequence approaches. Note that swatch frequencies for any radius $r$, and therefore the distance $d$, may be extended to such distributions: the frequencies are the probabilities that swatches of radius $r$ appear at the root of a random cell complex drawn from $\sigma$. This allows the distance from a finite cell complex to $\sigma$ to be computed, and makes the notion of a sequence of cell complexes converging to the probability distribution $\sigma$ sensible.

The limit distribution $\sigma$ is constructed by assigning probabilities to certain subsets of $\mathcal{C}^\bullet$ defined by swatches. Suppose that $S$ is a swatch of radius $r$, and $E_S$ is the set of all cell complexes in $\mathcal{C}^\bullet$ where $S$ appears at the root vertex. The probability of $E_S$ is then the limiting value of the swatch frequency of $S$ in the sequence $C_i$ as $i\rightarrow \infty$. It is a mathematical theorem \cite{2012lovasz} that this is sufficient to define the probability distribution $\sigma$ on the entire space $\mathcal{C}^\bullet$. By construction, $C_i$ converges to $\sigma$ in the sense that the distance $d$ between $C_i$ and $\sigma$ becomes arbitrarily small for sufficiently large $i$.

The convergence of a cell complex sequence to a limit implies the convergence of all local topological properties of that sequence as well. For example, the expected number of cycles of length four to which an edge belongs will converge. More precisely, let $H$ be the labeled adjacency graph of a square, let $G_i$ be the labeled adjacency graph of the cell complex $C_i$, and let $\text{inj}\paren{H,G_i}$ be the number of times $H$ appears in $G_i$. Although both $v\paren{G_i}$ (the number of vertices of $G_i$) and $\text{inj}\paren{H,G_i}$ will usually diverge with increasing $i$, if $\left\{G_i\right\}$ converges, the normalized quantity $\text{inj}\paren{H,G_i}/v\paren{G_i}$ will converge as well. 

More generally, the normalized number of adjacency preserving maps from $H$ to $G_i$ converges for any labeled graph $H$. This may be used to find, e.g., the probability that a 0-cell is adjacent to a specified number of 1-cells (as for the number of contacts around a sphere in an random sphere packing), the probability that a 1-cell is connected to 0-cells of specified degree (as with the Randi\`{c} index \cite{1975randic} of an organic molecule), the joint probability of adjacent 2-cells being incident to specified numbers of 1-cells (as in the Aboav-Weaire relation \cite{1974weaire,1980aboav} for a 2D microstructure), or the probability of a 1-cell participating in a cycle of specified length (as for ring statistics \cite{2010leroux} in an covalent glass). In this sense, the cloth provides a complete description of the local topology of the underlying cell complex, as initially claimed in Section \ref{sec_introduction}.

Consider a dynamical process on random cell complexex, and suppose that many of the properties of the system converge as time proceeds. The steady-state hypothesis is that there is a time interval when all scale-invariant properties of the network are constant in time, though this interval will depend on the initial conditions. To connect this to the formalism established above, construct a sequence of initial conditions of increasing size and allow all of them to evolve to the steady-state condition. By the steady-state hypothesis, the cloths of the systems will be identical up to finite size effects and the cell complex sequence will converge. This implies the existence of a universal limit distribution that may be viewed as a probability distribution of swatch types for an infinite steady state.

In practice, one can track the distance from a cell complex to a reference state as the cell complex evolves. If the steady-state hypothesis holds and the reference is in the steady state condition, then the distance will decrease toward zero and stabilize for a significant interval of time. Hence, the distance provides a powerful tool to test the steady-state hypothesis. In Section \ref{sec_grain_growth}, we use this to study the convergence properties of a model grain boundary network.

Finally, we note that the subject of convergent cell complex sequences is deeper than may be apparent from this section. For instance, the analysis of global graph properties of a convergent cell complex sequence (i.e., those that can be expressed as maps from the adjacency graph $G_n$ into a graph $H$) is much more difficult than that of local topological properties, and not all of them are convergent. An example of a convergent global graph property is the number $q$-colorings of $G_n$ for sufficiently high $q$, i.e., the number of different ways that $q$ colors may be assigned to the vertices of $G_n$ such that no vertices connected by an edge have the same color. A second point is that the swatch frequencies in a cell complex are far from independent---given a swatch of radius $r$, the swatches with roots on the neighboring vertices are determined up to radius $r-1$. This fact is reflected by an important property of the limit probability distribution called involution invariance. Further exposition of these subjects is beyond the scope of this paper; the interested reader is encouraged to refer to the book by Lovasz \cite{2012lovasz}.

\section{Two Models of Grain Growth}
\label{sec_grain_growth}

Microstructure evolution in polycrystalline materials is quite complicated, with the general case involving the precipitation of solid phases, the diffusion of solute species, the formation of dislocation networks, and the interaction of stress fields with all of the above processes. Along with the scarcity of experimental values for many of the relevant material properties, this means that practical simulations of microstructure evolution often require a number of simplifying assumptions.

Perhaps the simplest system with a nontrivial evolution is a pure polycrystalline material with negligible stored strain energy. This system is represented in two dimensions by a space-filling set of grains, with two adjacent grains meeting on a grain boundary and three adjacent grain boundaries meeting on a triple junction. The disruption of the atomic bonding along the boundaries endows them with an energy per unit length, and the minimization of this energy drives the motion of the boundaries and a concomitant increase in the average area of the grains. Hence, grain growth is a result of boundary motion, and boundary motion is usually described by the Turnbull relation \cite{1951turnbull}
\begin{equation}
v_n = m_0 \exp \! \left( -\frac{Q_{gb}}{k_B T} \right) p. \nonumber
\end{equation}
Here $v_n$ is the boundary velocity in the normal direction, $m_0$ is the mobility prefactor, $Q_{gb}$ is the activation energy for boundary motion, and $p$ is the driving pressure. For a pure polycrystalline material with negligible strain energy and a constant boundary energy per unit length $\gamma$ \cite{1992taylor}, the pressure on a boundary is given by the Young--Laplace equation \cite{1805laplace}
\begin{equation}
p = \gamma \kappa,
\end{equation}
where $\kappa$ is the boundary curvature. Further assuming a constant boundary mobility $m$ allows the governing equation to be reduced to
\begin{equation}
v_n = m \gamma \kappa,
\label{eq_curvature}
\end{equation}
or the equation for curvature driven motion. This is the starting point for most simulations of grain growth in two dimensions.

\begin{figure}
\center
\subfloat[]{%
	\label{fig_node_continuous}{%
		\includegraphics[width=2.5cm]{%
			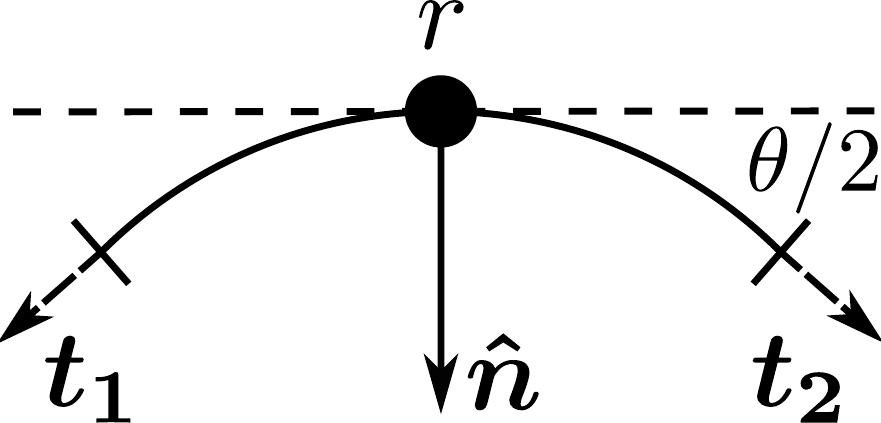}}}
\hspace{8pt}
\subfloat[]{%
	\label{fig_node}{%
		\includegraphics[width=2.5cm]{%
			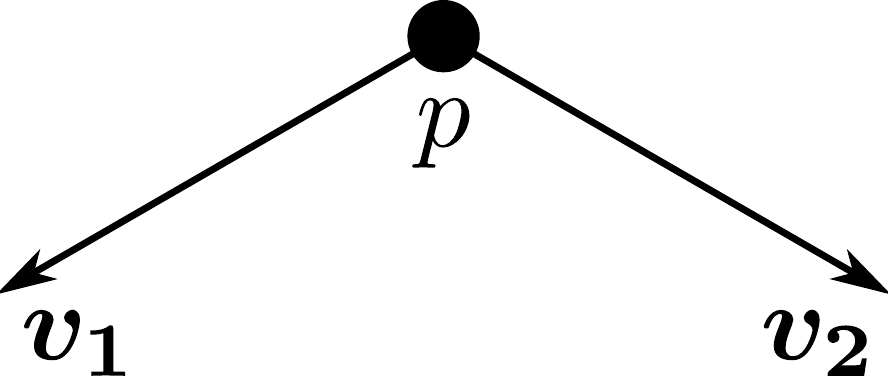}}}
\hspace{8pt}
\subfloat[]{%
	\label{fig_vertex}{%
		\includegraphics[width=2.5cm]{%
			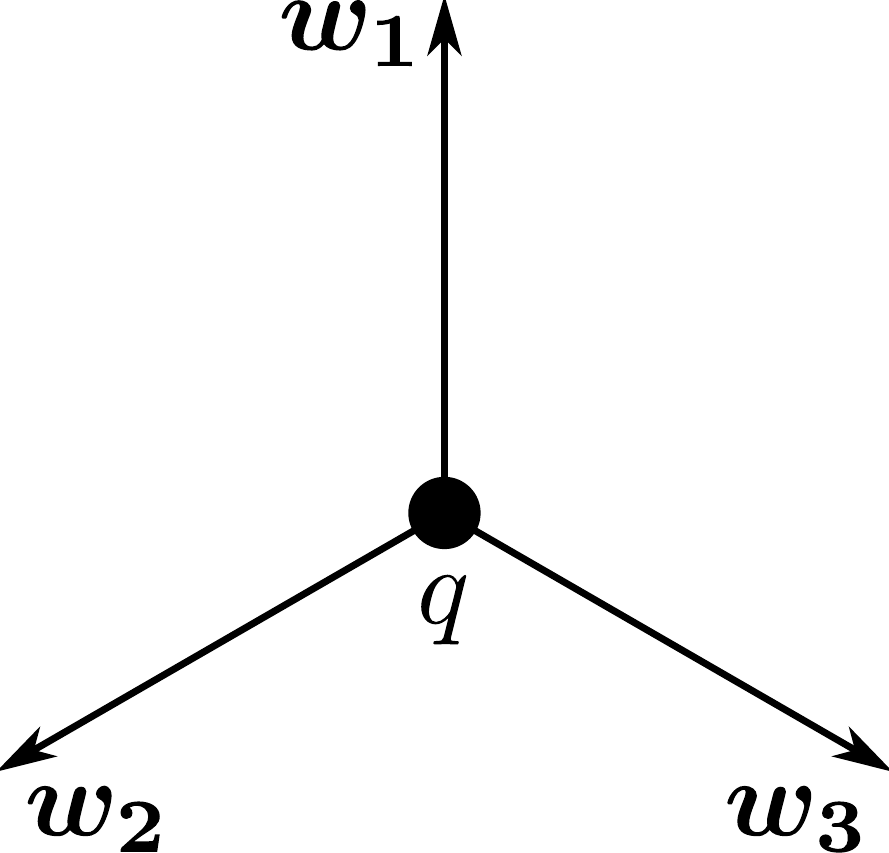}}}
\caption{\label{transitions} (a) A continuous grain boundary edge. (b) Node $p$ joins two discrete boundary edges. (c) Vertex $q$ joins three discrete boundary edges.}
\end{figure}

The microstructure is modeled as a network of polygonal curves in two dimensions. Grain boundaries consist of line segments that meet at nodes of degree two, and triple junctions consist of vertices of degree three. The nodes and vertices of the grain boundary network correspond to the node $p$ and the vertex $q$ of Figure \ref{transitions}, respectively. Equations of motion for the nodes are derived by considering the boundary edge in Figure \ref{fig_node_continuous}, and assuming that this edge is a planar curve of length $\Delta s$. If the length is sufficiently small, then the curvature is effectively constant and the edge can be considered as an arc of a circle. Let the angle subtended by the arc be $\Delta \theta$, the point halfway along the arc be $r$, and the normal vector at $r$ be $\nvec{n}$. The force on this edge arises from the line tension $\gamma$ being applied to the segment endpoints along the tangent vectors $\nvec{t}_1$ and $\nvec{t}_2$, and the force per unit length of boundary is $\gamma (\nvec{t}_1+\nvec{t}_2) / \Delta s$. Multiplying the force per unit length by the boundary mobility $m$ gives
\begin{align}
\frac{\Delta \vec{r}}{\Delta t} &= \frac{m \gamma (\nvec{t}_1+\nvec{t}_2)}{\Delta s} = m \gamma \frac{2 \sin(\Delta \theta / 2)}{\Delta s} \nvec{n} \nonumber \\
&\approx m \gamma \kappa \nvec{n} \nonumber
\end{align}
for the velocity of $r$, where the displacement $\Delta \vec{r}$ occurs in a time interval $\Delta t$. The second equality follows from $\nvec{t}_1$ and $\nvec{t}_2$ having equal and opposite components perpendicular to $\nvec{n}$, and components of length $\sin(\Delta \theta / 2)$ parallel to $\nvec{n}$. The approximate equality uses the small angle sine approximation and defines the curvature as $\kappa = \Delta \theta / \Delta s$. This is precisely the equation of curvature driven motion, and suggests that the equation of motion for the nodes of the discrete case can be derived in a similar manner.

The configuration in Figure \ref{fig_node} is the discrete version of the continuous boundary edge in Figure \ref{fig_node_continuous}. Two adjacent segments intersect at the node $p$, and the vectors $\vec{v}_1$ and $\vec{v}_2$ extend from $p$ to the two adjacent nodes. An equitable partition of the grain boundary network assigns half of the segments along $\vec{v}_1$ and $\vec{v}_2$ to $p$, and the remaining half to the adjacent nodes. The node $p$ is therefore associated with a boundary length of $(\norm{\vec{v}_1} + \norm{\vec{v}_2}) / 2$. The force that arises from the line tension $\gamma$ being applied along the vectors $\vec{v}_1$ and $\vec{v}_2$ is $\gamma(\nvec{v}_1+\nvec{v}_2)$, where $\nvec{v}_i$ is the unit vector along $\vec{v}_i$. As before, multiplying the force per unit length by the boundary mobility $m$ gives
\begin{equation}
\frac{\Delta \vec{p}}{ \Delta t} = \frac{2 m \gamma (\nvec{v}_1+\nvec{v}_2)}{\norm{\vec{v}_1} + \norm{\vec{v}_2}}
\label{eq_node}
\end{equation}
for the velocity of $p$, where the displacement $\Delta \vec{p}$ occurs in a time interval $\Delta t$. This is a suitable discrete approximation for curvature driven motion provided that the small angle sine approximation holds. That is, the exterior angle in Figure \ref{fig_node} should be small. Our simulations satisfy this condition by dynamically interpolating the polygonal curves to keep the exterior angle at every node below $\pi/10$.

Notice the absence of an equation of motion for triple junctions in the continuous system. The reason for this is that requiring the boundaries to move by curvature flow does not uniquely specify the behavior of the triple junctions, though any equation of motion must satisfy the following physical constraint. Let $M$ be a mobility of the triple junctions that is distinct in units and value from the mobility $m$ of the boundaries. Any equation of motion should cause the angles between boundary edges in an infinitesimal neighborhood around a triple junction to approach $2\pi/3$ in the limit of high $M.$ Conversely, decreasing $M$ should increase the deviation of the angles from $2\pi/3$ for a triple junction subject to a constant driving force.

A simple equation of motion for the triple junctions that is consistent with the above physical constraint is given by the following line of reasoning. With reference to Figure \ref{fig_vertex}, the force on the triple junction $q$ that arises from the line tension $\gamma$ being applied along the vectors $\vec{w}_1$, $\vec{w}_2$ and $\vec{w}_3$ is written as $\gamma (\nvec{w}_1 + \nvec{w}_2 + \nvec{w}_3)$, where $\nvec{w}_i$ is the unit vector along $\vec{w}_i$. Multiplying this force by the triple junction mobility $M$ gives
\begin{equation}
\frac{\Delta \vec{q}}{\Delta t} = M \gamma (\nvec{w}_1 + \nvec{w}_2 + \nvec{w}_3)
\label{eq_vert}
\end{equation}
for the velocity of $q$, where the displacement $\Delta \vec{q}$ occurs in a time interval $\Delta t$. Since Equation \ref{eq_vert} assigns a finite mobility $M$ to the vertices, Equations \ref{eq_node} and \ref{eq_vert} will be called the finite mobility equations. Note that there is some evidence of finite triple junction mobilities in the literature \cite{2000gottstein,2002gottstein}.

The finite mobility equations are not the only option. Instead, one can require that the grain boundaries meet at angles of $2 \pi / 3$ (called the Herring Angle condition), which provides a reasonable set of boundary conditions for the curvature flow along the edges.~\cite{2001kinderlehrer,2014ilmanen} This can be interpreted as giving the vertices infinite mobility, as they always move to the point which locally minimizes the lengths of the neighboring edges. The Herring Angle condition also implies that the rate of change of the area $A$ of a grain with $n$ bounding triple junctions is given by the von Neumann--Mullins relation \cite{1952vonneumann,1956mullins}
\begin{equation}
\deriv{A}{t} = m \gamma \frac{\pi}{3} ( n - 6 ).
\label{eq_mullins}
\end{equation}
That is, the rate of change of the area depends linearly on the number of bounding triple junctions. While Equation \ref{eq_mullins} is a consequence of Equation \ref{eq_curvature} and the Herring Angle condition, the quantities appearing in Equation \ref{eq_mullins} can be more reliably measured than the normal direction and curvature of a polygonal curve. This motivated the development of an alternate set of equations of motion based on the von Neumann--Mullins relation \cite{2010lazar} that will be called the von Neumann--Mullins equations.

\begin{figure}
\center
\subfloat[]{%
	\label{move_1}{%
		\includegraphics[height=1.25cm]{%
			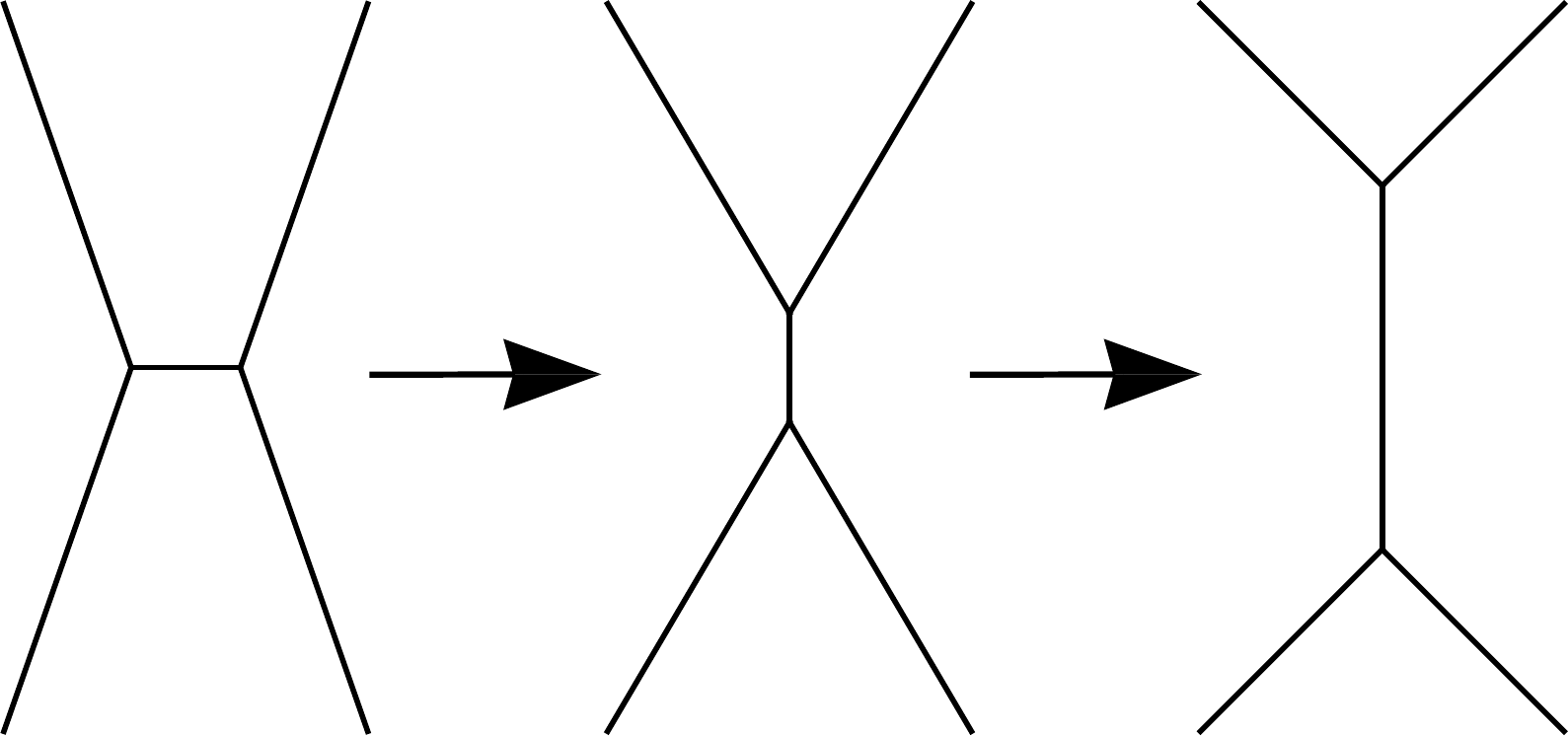}}}
\hspace{24pt}
\subfloat[]{%
	\label{move_2}{%
		\includegraphics[height=1.25cm]{%
			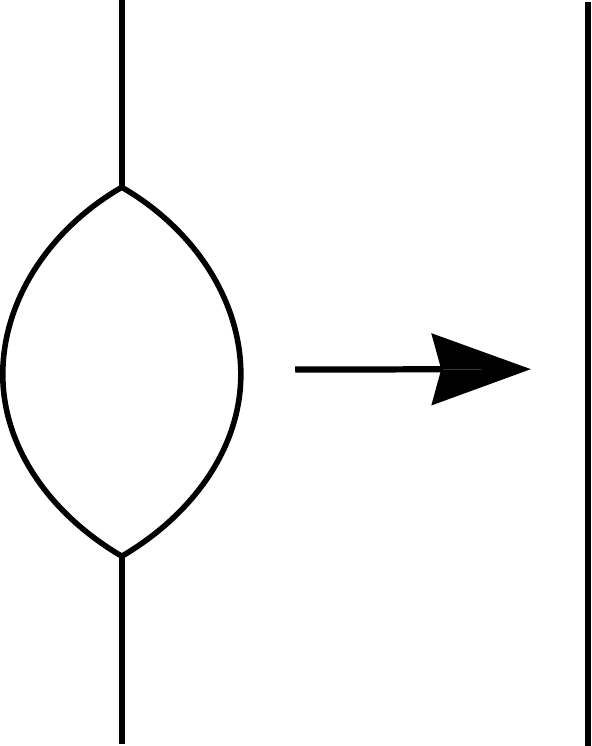}}}
\caption{\label{moves} Topological operations that occur in the grain growth simulations include (a) edge flips and (b) digon deletions.}
\end{figure}

Two topological operations are allowed in the simulation, namely, the flip of an boundary edge and the deletion of a digon. A boundary flip occurs whenever the length of a boundary passes below a threshold value and the creation of a degree-four vertex appears imminent, as in Figure \ref{move_1}. The boundary direction is changed and connections with adjacent boundaries are shuffled to minimize the sum of the two angles opposite to the flipped boundary. A digon is deleted whenever the maximum distance between points on the two boundaries passes below a threshold value, as in Figure \ref{move_2}. One of the boundaries is deleted, and the remaining boundary is merged with the two adjacent boundaries. Note that the topological change that occurs when a grain shrinks to a point can be expressed as a combination of the two previous operations.

\begin{figure}
\center
\subfloat[]{%
	\label{initial_grain}{%
		\includegraphics[width=3.9cm]{%
			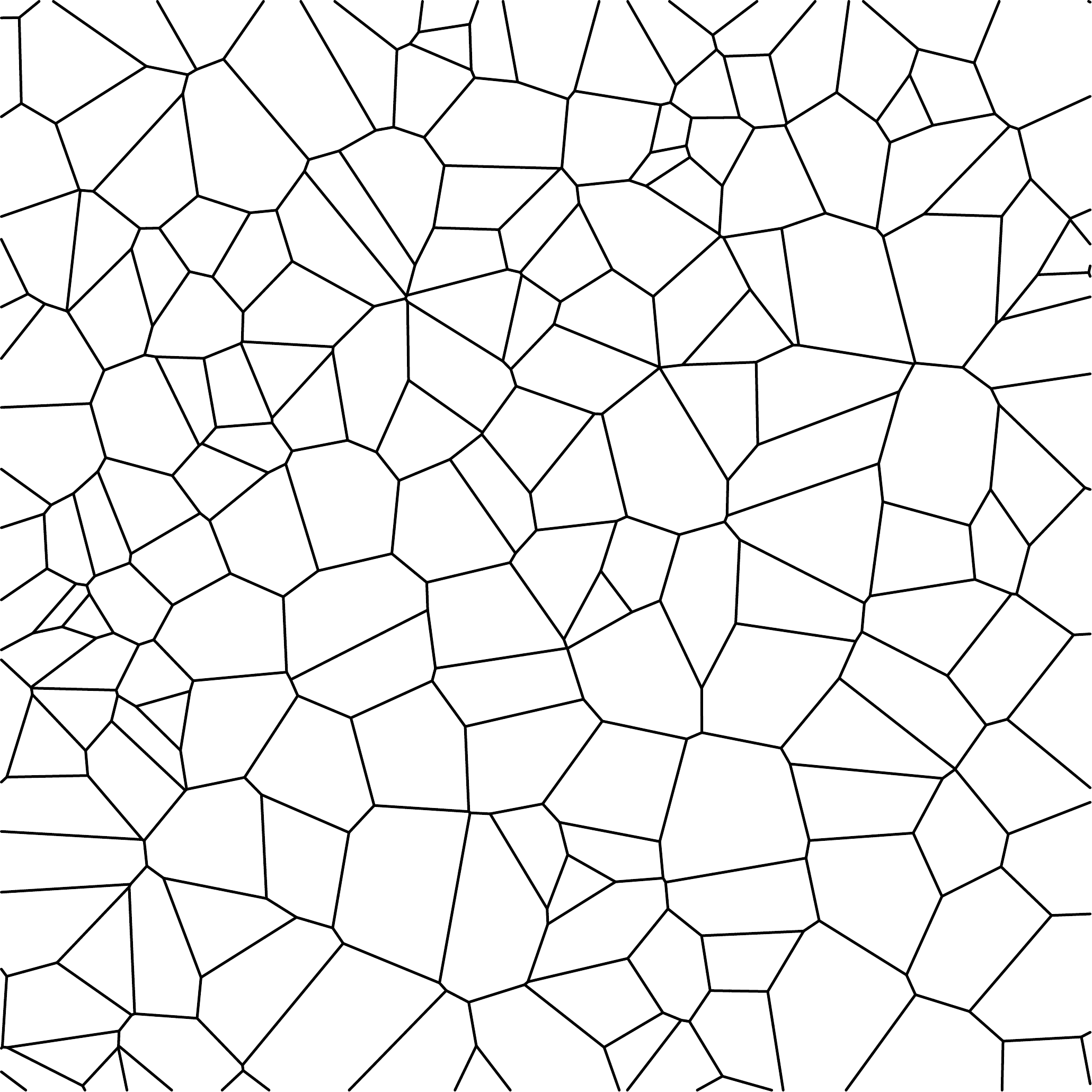}}}
\hspace{14pt}
\subfloat[]{%
	\label{perturbed_lattice}{%
		\includegraphics[width=3.9cm]{%
			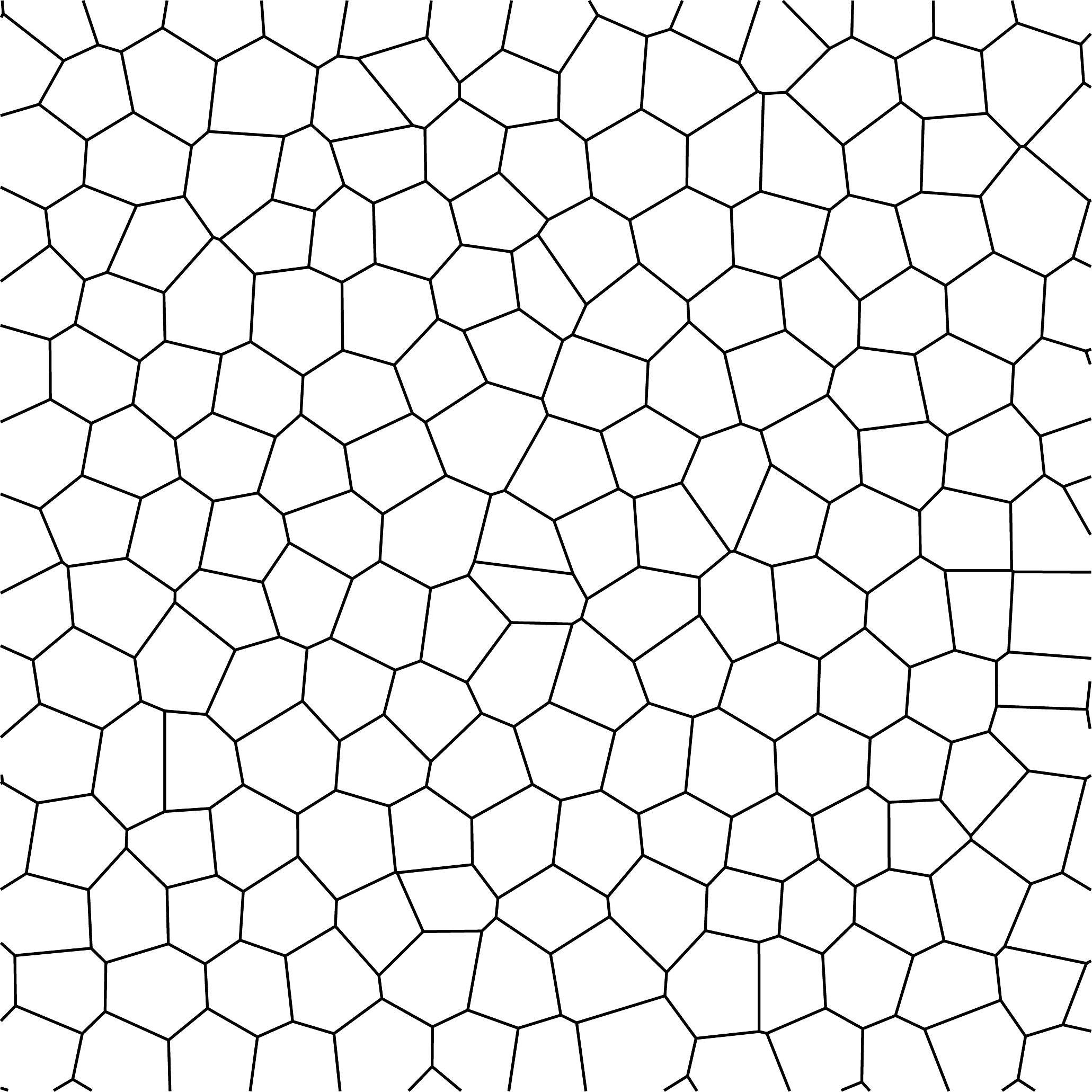}}}
\caption{\label{grain_growth}Initial conditions for the grain boundary simulation: (a) a Voronoi tessellation of randomly distributed points and (b) a perturbed lattice.}
\end{figure}

Grain growth simulations were always performed in a square with periodic boundary conditions, with two types of initial conditions. The first is given by the Voronoi diagram of randomly distributed points, and a portion of this initial condition is shown in Figure \ref{initial_grain}. The second is a perturbed honeycomb lattice, and a portion of this is shown in \ref{perturbed_lattice}. It is generated by perturbing the vertices of the dual triangular lattice with displacements independently sampled from a two-dimensional Gaussian distribution with a standard deviation of one-fourth the lattice spacing, then computing the Voronoi diagram.

\section{A Statistical Steady State}
\label{sec_steady_state}

We used one of the cell complex distances to compare the model grain boundary networks to a reference condition throughout the simulations. The reference condition was a network with $3.1 \times 10^6$ boundaries resulting from a simulation that used the von Neumann--Mullins equations, that began from a Voronoi tessellation with $6.0 \times 10^7$ boundaries, and for which all measured scale-free properties had  converged. Since the number of swatch types increases dramatically as a function of radius, the sample size required to accurately compute the cloth increases dramatically as well. Practically, $r = 7$ was the largest radius that gave reliable cloth statistics; an appreciable number of swatch types occurred only once for $r = 8$ in all of our simulations, indicating that larger samples would be needed. Nevertheless, there are so many swatch types for this radius that the cloth still offered a very detailed description, and we use the distance $d_r$ with $r = 7$ instead of the limit distance $d$ in the following.

\begin{figure}
\includegraphics[height=5.8cm]{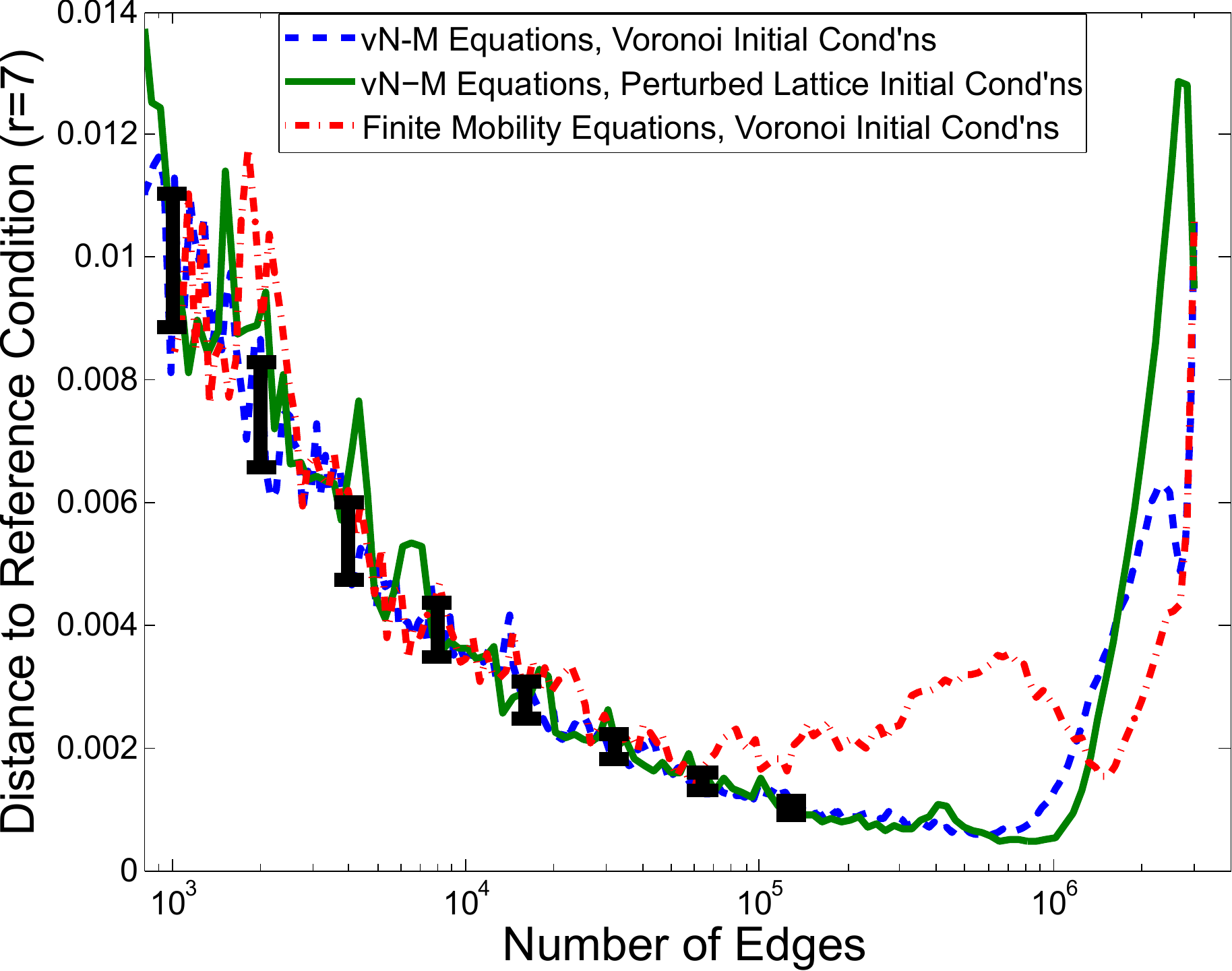}
\caption{\label{bootstrap}Distances of several simulations to the reference condition. Error bars show the standard deviation of the distance of a steady-state configuration with the indicated number of edges to the reference condition.}
\end{figure}

\begin{figure}
\includegraphics[width=3.9cm]{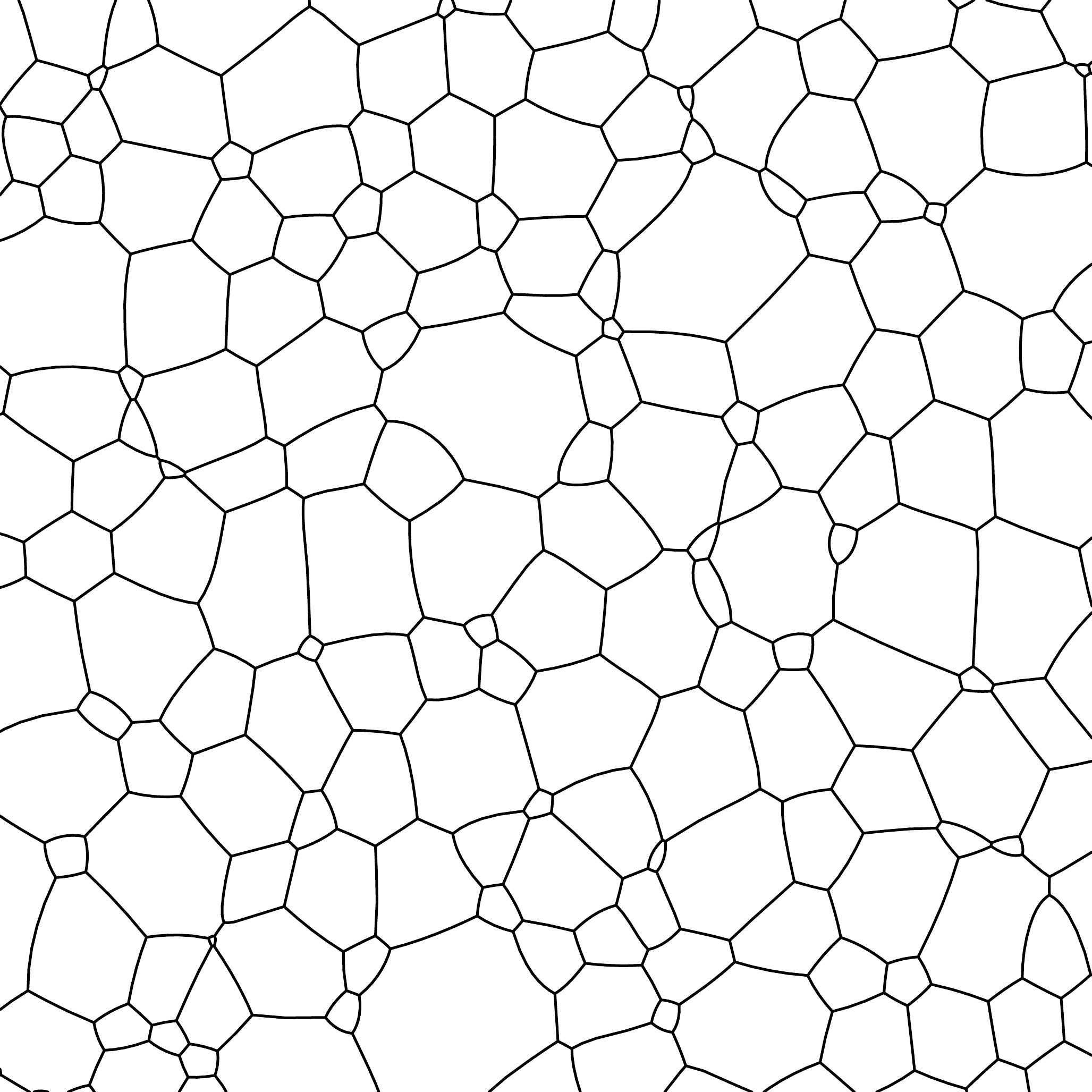}
\caption{\label{mullins}The steady-state condition of the model grain boundary network.}
\end{figure}

Figure \ref{bootstrap} shows the distance to the reference condition for three simulations, one using the finite mobility equations and the other two the von Neumann--Mullins equations. For the latter two, the distance to the reference condition decreased very rapidly as they evolved, indicating convergence toward the steady state depicted in Figure \ref{mullins}. The simulation using the finite mobility equations also approached the reference condition, but not as quickly. The evolution of the systems is parametrized by the number of edges, a non-increasing function of time.

While Figure \ref{bootstrap} indicates that the distance to the reference condition initially decreases, the distance to the reference condition visibly increases for small numbers of remaining edges. This may be explained by the decrease in the sample size increasing the statistical error in the swatch frequencies and the apparent distance to the reference condition. A test of convergence to the steady state should account for this source of error. Let $R_n$ be a set of networks with $n$ edges that are already in the steady state. A measured distance may be compared to the distribution of distances from elements of $R_n$ to the reference condition; if the measured distance falls within one standard deviation of the mean of this distribution for a long time interval, then the network being considered is likely in the steady state. Practically, the set $R_n$ contains random subsamples of the reference condition. A single subsample is constructed from the vertices and edges within some radius of a randomly selected vertex, with vertices on the boundary excluded as necessary to attain the desired number of edges.

This procedure is used to evaluate the convergence of the simulations in Figure \ref{bootstrap}. The simulations began with $3.0 \times 10^7$ boundary edges. The error bars in the figure extend one standard deviation above and below the mean distance from the elements of $R_n$ to the reference condition. Note that the simulations using the von Neumann--Mullins equations (the solid green line and the evenly dashed blue line) are within one standard deviation of the subsamples throughout the interval between $100,000$ and $1000$ edges, and that the same is true for the simulation using the finite mobility equations (the dashed line) between $50,000$ and $1000$ edges. This offers strong evidence that they have converged to the steady state in the indicated intervals. Further evidence in the form of various statistical quantities suggests that the simulations using the von Neumann--Mullins equations converged with as many as $500,000$ edges remaining, though the reference condition does not contain enough edges to allow independent subsamples of that size. This data leads to a conjecture:
\begin{conj}
There exists a unique limit distribution $\sigma$ on the space of all countable, connected one-dimensional cell complexes with a root cell specified such that all generic initial conditions converge to $\sigma$ under the von Neumann--Mullins Equations.
\end{conj}
Of course, a definition of ``generic initial conditions'' is required for this to be a mathematical precise conjecture. This turns out to be a delicate question, and is addressed in a separate paper for a mathematical audience~\cite{2015Schweinhartb}.
 
\begin{figure}
\includegraphics[height=5.8cm]{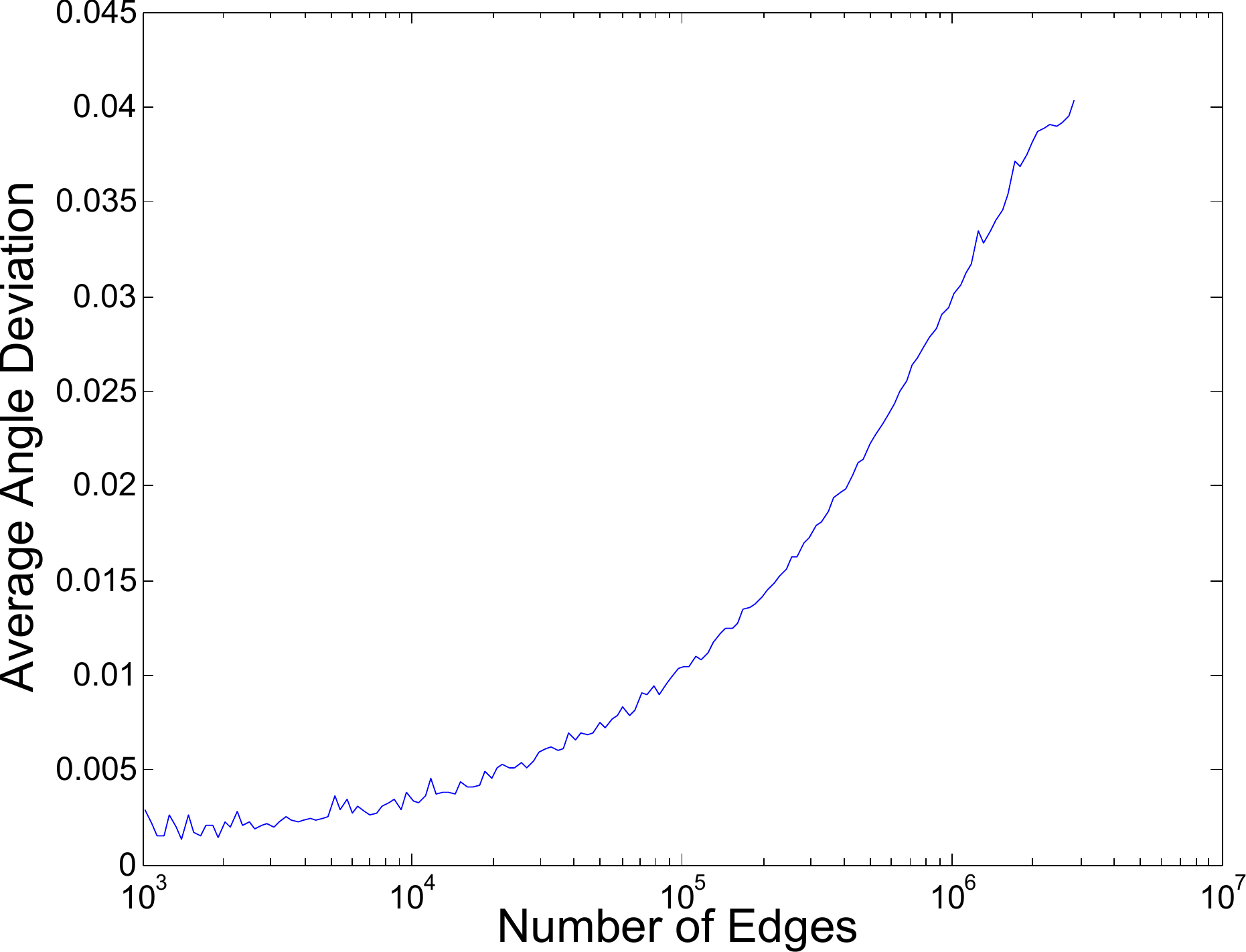}
\caption{\label{fig_angles} The average deviation of the triple point angles from $2\pi/3$ for a simulation using the finite mobility equations.}
\end{figure}

It is not obvious \textit{a priori} that the finite mobility equations should approach the same steady state as the von Neumann--Mullins Equations. A possible explanation is that is a consequence of the fact that curvature is not a scale-free property: a circle of half the radius has twice the curvature. Thus, as the system coarsens and the average edge length increases, the average speed of an edge will decrease. In contrast, the finite vertex mobility equations are scale invariant. As a result, the vertices move faster and faster relative to the edges, in effect causing their relative mobility to increase toward infinity. It follows that the triple point angles should approach $2\pi/3,$ and therefore that the system should behave more like the von Neumann--Mullins Equations. This is corroborated by Figure \ref{fig_angles} which shows how the average deviation of the triple point angles from $2\pi/3$ changes in a simulation using the finite vertex mobility equations. The simulation is the same finite mobility system from Figure \ref{bootstrap}.

This reasoning also explains why the convergence is slower for the finite mobility equations, as they only drift toward the infinite mobility case. Still, they provide a good approximation of this behavior in the long term, which will be important for the 3D case where (as far as we know) there is no discrete, physical way to directly simulate a system with infinite vertex mobility.  Let us note that a family of different universal conditions can be obtained by rescaling the mobility of the vertices as the simulation evolves to keep their speed relative to the edges constant. The resulting behavior is mathematically interesting, and is discussed in length in B. Schweinhart's thesis \cite{2015schweinhart, 2015Schweinhartb}.

\section{A Model Dislocation Network}
\label{sec_dislocations}

This section applies the concepts of a swatch and a cloth to measure the approach of a model dislocation network to a topological steady state. More specifically, the dislocation network is modeled as a network of polygonal curves in three dimensions. Line segments composing a dislocation meet at nodes of degree two, while dislocations meet at vertices of degree three. The complex calculations required to model the interactions of dislocation edges are neglected, and only the dislocation self energy is retained. That is, a dislocation edge is given a constant energy $\gamma$ per unit length and evolves following a simple line-tension model \cite{1982hirth}. There is evidence in the literature \cite{1967lothe,1967indenbom} that this approach is justified when considering the general characteristics of a complex dislocation network rather than specific dislocation reactions. Since our purpose is to study the statistical topology of a dislocation network instead of the effect of dislocations on material properties, we believe that this is a reasonable simplification. This system is also of intrinsic mathematical interest.

\begin{figure}
\center
\includegraphics[height=1.34cm]{%
	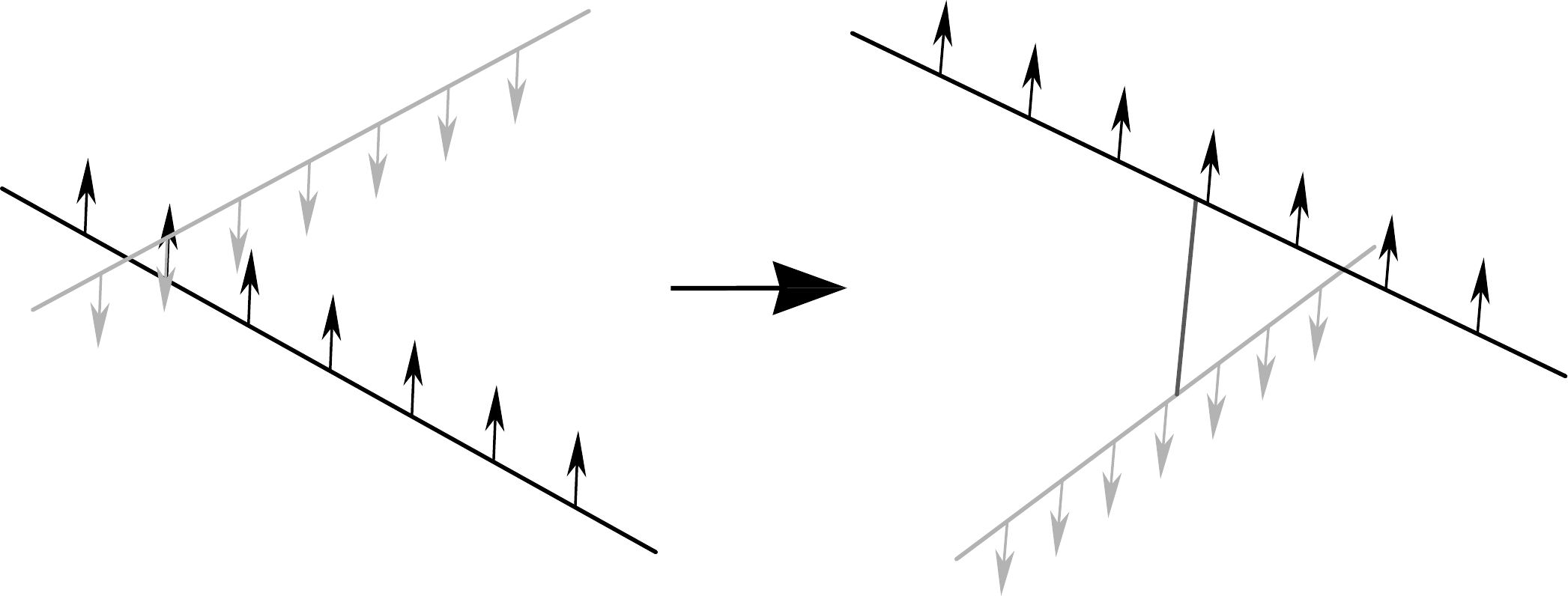}
\caption{\label{move_3}The third topological operation that occurs in the dislocation simulations is the edge intersection, resulting in the joining of edges.}
\end{figure}

The three topological operations that are allowed in the simulation include the two analogues of the operations in Figure \ref{moves} (an edge flip and a digon deletion) and the edge intersection, shown in Figure \ref{move_3}. An edge intersection occurs whenever two non-neighboring edges meet transversely. The edges are subdivided at the point of intersection and joined by a newly created edge. Since detecting intersections is computationally expensive, the simulation could be made substantially more efficient if this topological operation could be neglected without measurably changing the topological steady state. Using the metric on cloths introduced in Section \ref{sec_metric}, we provide evidence that this is indeed the case.

\begin{figure}
\center
\subfloat[]{%
	\label{vor}{%
		\includegraphics[width=3.9cm]{%
			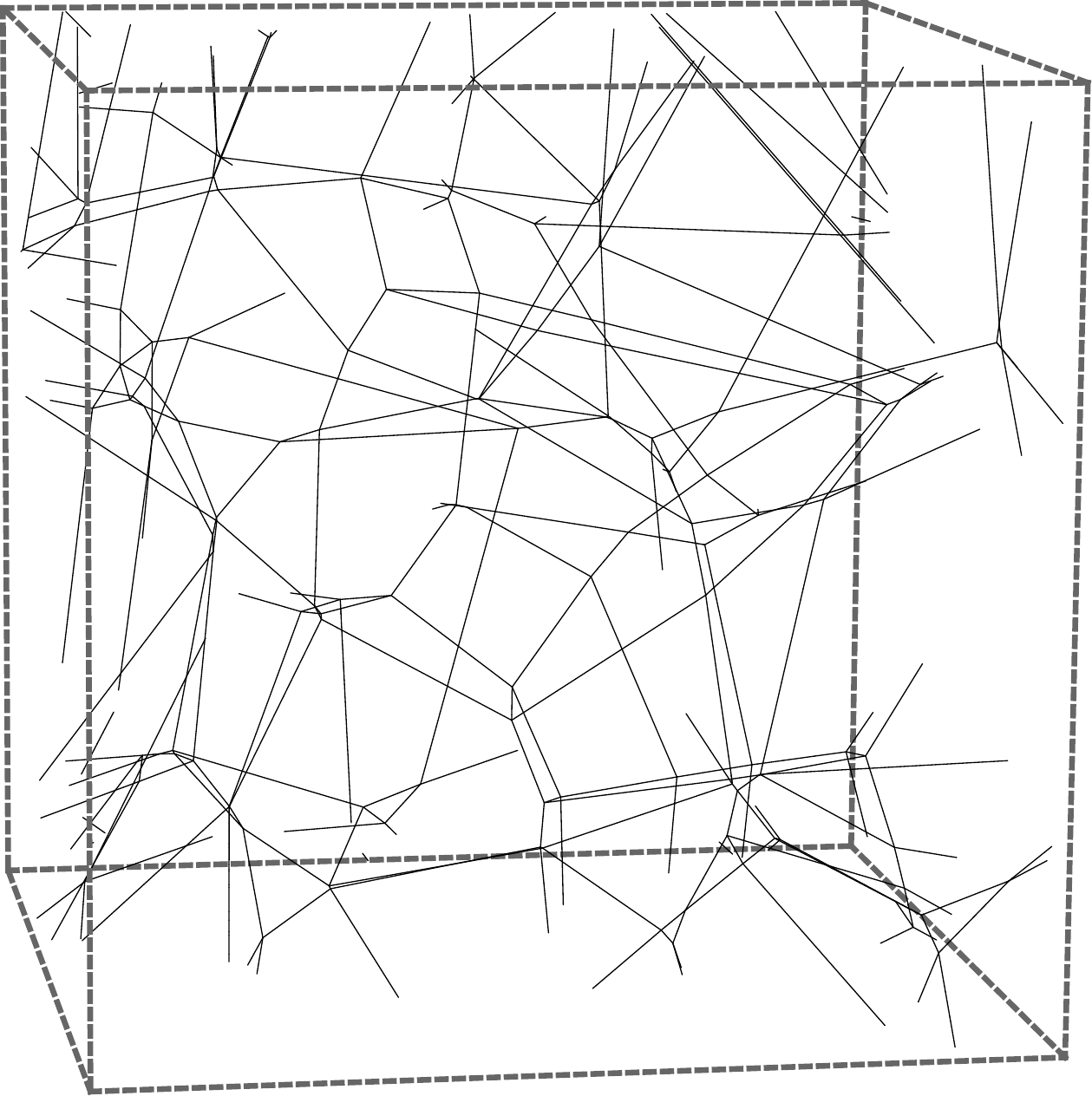}}}
\qquad
\subfloat[]{%
	\label{rand}{%
		\includegraphics[width=3.9cm]{%
			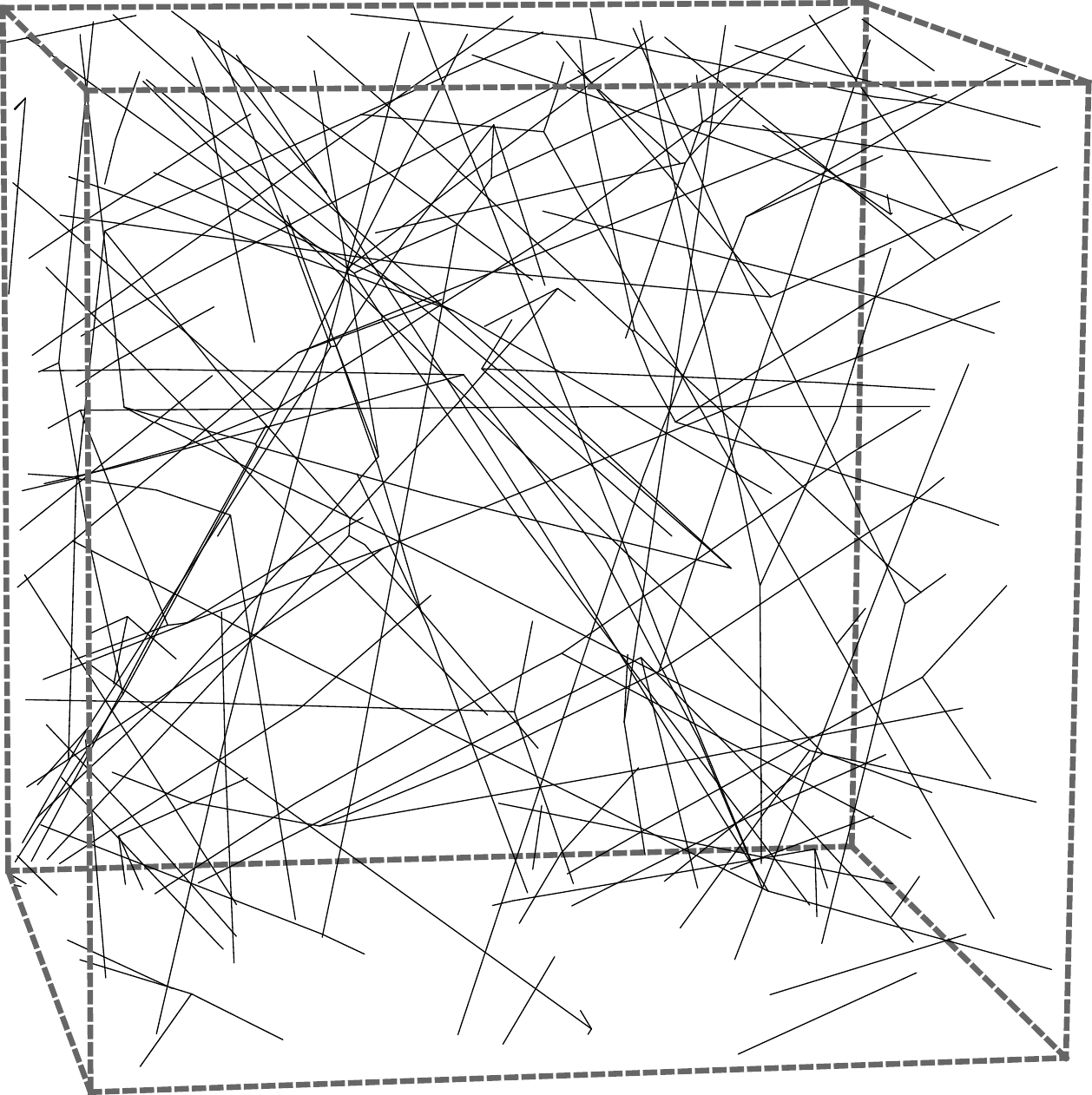}}}
\caption{\label{initial}Initial conditions of the dislocation network included (a) Voronoi graphs and (b) random graphs, with the constructions given in the text. Note that (b) shows a smaller volume than (a), since edges are much denser in the random graph.}
\end{figure}

The simulations were performed in a cube with periodic boundary conditions. Initial conditions for the simulations were generated from a set of random points by one of two procedures. The first was constructed from the edge set of the Voronoi diagram of the points. Since the vertices of this network had degree four, the network had to be modified to make an admissible dislocation network. Every vertex was replaced by an edge, and the four adjacent edges were assigned to the vertices to minimize the maximum angles opposite to the newly created edge. The resulting initial condition is called a Voronoi graph, and is depicted in Figure \ref{vor}. Ken Brakke's VOR3DSIM program was used to compute Voronoi tessellations \cite{BrakkeVor}. For the second procedure, points of degree less than three were randomly connected by edges with others closer than a threshold distance. If there were no points within the threshold, two cases were considered. If the point had degree two, it was considered to be a node along the edge between its two neighbors. Otherwise, it was paired with the closest possible point. This process proceeded until the creation of edges was no longer possible. The resulting initial condition is called a random graph, and is depicted in Figure \ref{rand}.

Since the model grain boundary network and the model dislocation network both satisfy the same conditions at the nodes and vertices and evolve by curvature-driven motion, Equations \ref{eq_node} and \ref{eq_vert} (the finite mobility equations) can also be used as equations of motion for the model dislocation network.

\begin{figure}
\includegraphics[height=5.8cm]{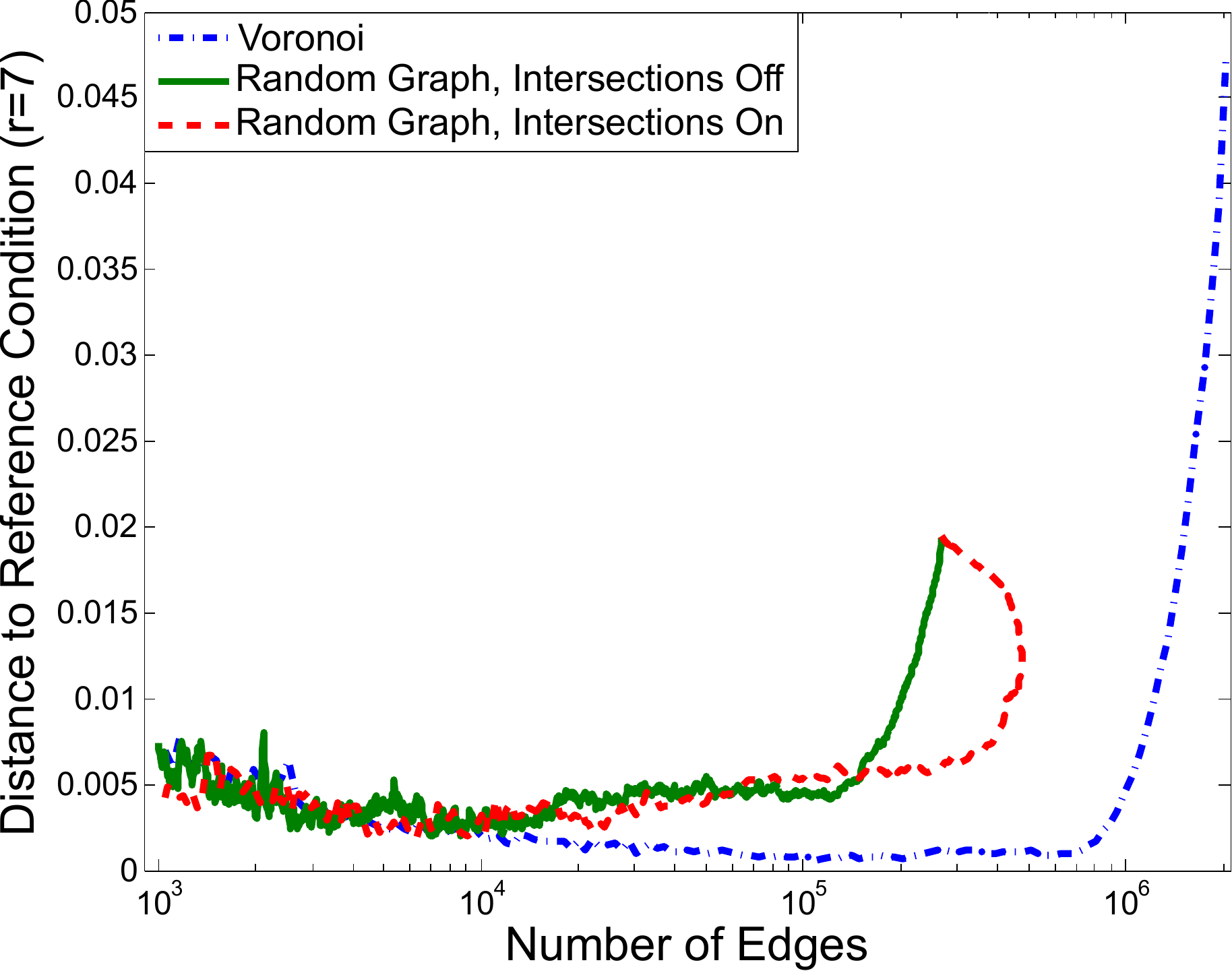}
\caption{\label{convergence} Distances of several simulations to the reference condition.}
\end{figure}

\begin{figure}
\includegraphics[width=3.9cm]{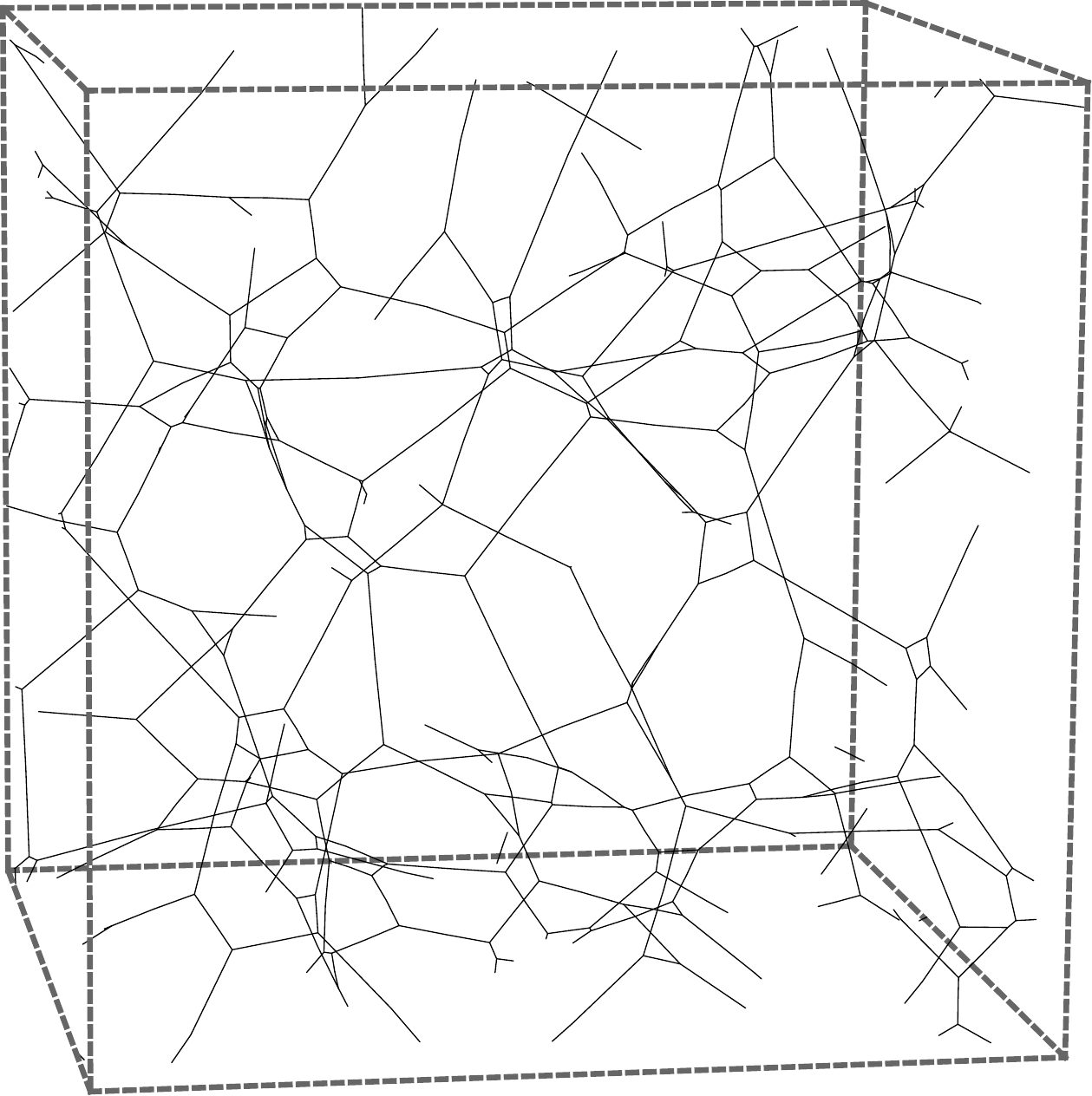}
\caption{\label{steadystate} A small region in the steady-state condition for the model dislocation network.}
\end{figure}

\begin{figure}
\includegraphics[height=5.8cm]{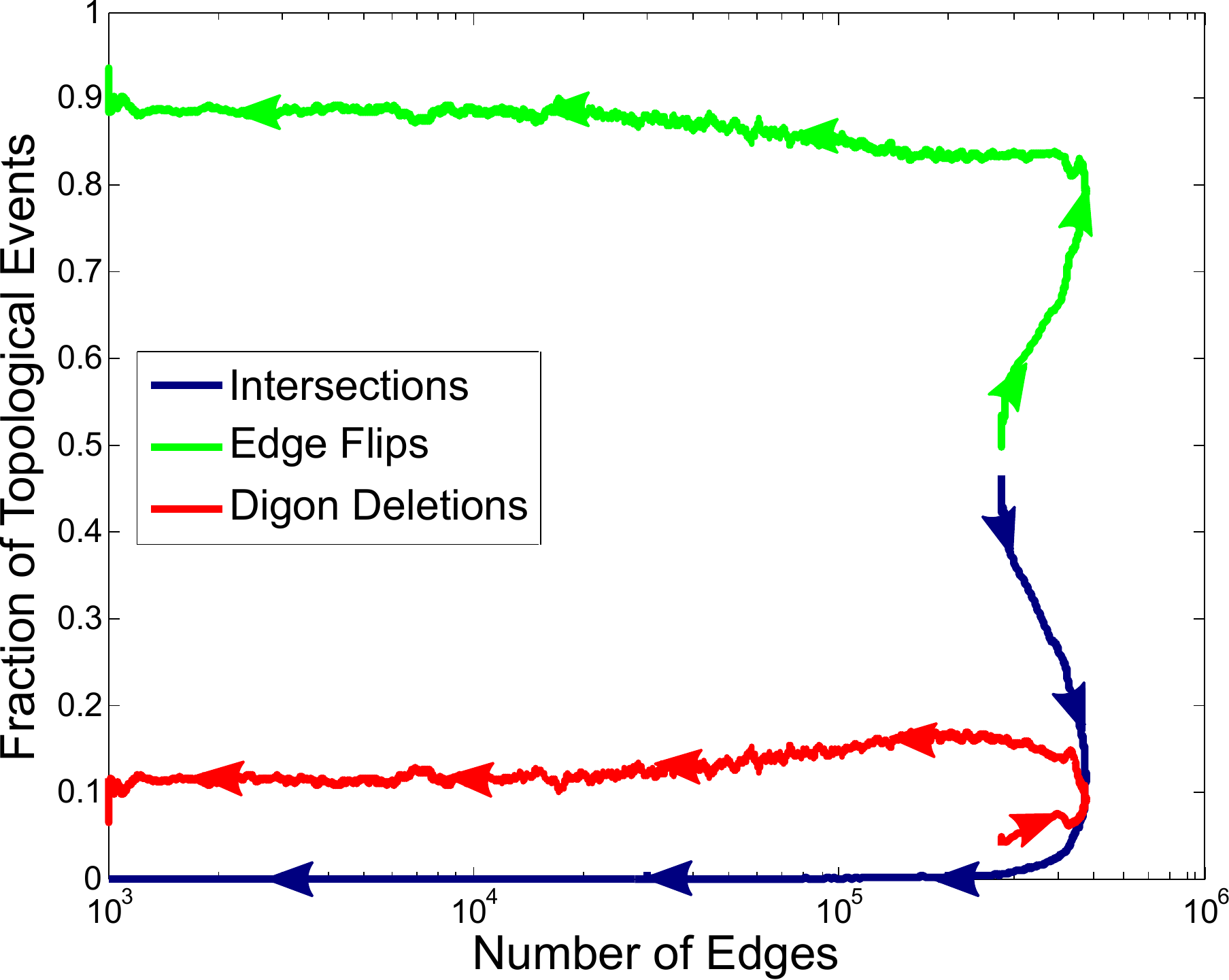}
\caption{\label{intersections} Relative rates of topological changes throughout a simulation with a random graph initial condition.}
\end{figure}

To test the convergence of the simulations, we tracked their distance to a reference condition throughout their evolution. The reference condition resulted from a simulation with $\lambda=1667$ starting with a Voronoi graph with slightly more than $10^7$ edges, and has about $10^6$ edges. As discussed in the final paragraph of the previous section, the finite mobility equations are not scale invariant, so a system evolving by them can only drift toward a topological steady state. Thus, we must be careful when selecting a reference condition, and comparing other simulations to it. However, if the mobility and the number of initial edges are large enough, the system should be very close to a steady state condition toward the end of its evolution. Our measurements of several properties support this, and it appears that any deviation from the topological steady state due to the finite value of the mobility is small relative to the statistical variation due to finite size effects. 

Figure \ref{convergence} shows the distance to the reference condition for three simulations. For all three, the distance to the reference condition decreased rapidly as they evolved, indicating convergence toward the state depicted in Figure \ref{steadystate}.  The evolution of the systems is parametrized by the number of edges in the system, which generally decreases with time but can increase if there are frequent edge intersections. For example, the simulation for one of the random graph initial conditions (shown by the dashed red line in Figure \ref{convergence}) initially experienced many intersections, leading to a transient where the number of edges increased. As shown in Figure \ref{intersections} though, the number of intersections as a fraction of all topological changes declined as the simulation progresses, eventually decreasing to almost none in the long term. This suggests that the long-term behavior may be insensitive to the detection of edge intersections. As further evidence for this, the remaining two simulations in Figure \ref{convergence} did not detect edge intersections, and yet converged toward the same state.

\section{Conclusion}
\label{sec_conclusion}

Although random cell complexes occur throughout the physical sciences, our ability to characterize them appears to have been limited by the absence of a suitable language. This paper proposes that the topology of the cell complex be represented by a graph. A swatch (defined in Section \ref{sec_swatches}) characterizes the local topology of the cell complex around a root vertex, and provides a description of the local environment that is agnostic to the details of the physical system. A cloth (defined in Section \ref{sec_cloths}) is constructed from the probabilities of every swatch type occurring around a randomly selected root vertex, and may be used to answer any question pertaining to the statistical local topology of the cell complex. This includes, e.g., the distribution of the number of contacts around a sphere in a random sphere packing, the distribution of the number of sides in the rings in a covalent glass, and the distribution of the number of faces in the grains of a polycrystal.

For cell complexes that evolve by some dynamical process, a sequence of cell complexes can be constructed at successive points in the evolution. A distance on cell complexes is defined in Section \ref{sec_metric} such that the elements of this sequence become arbitrarily close together if the system evolves toward a steady state. This allows a precise definition of the limiting condition to be given. This was applied to a two-dimensional grain boundary network with uniform boundary energies and mobilities, and obeying one of two different sets of equations of motion. The first assumed an infinite vertex mobility, and is designed \cite{2010lazar} to accurately satisfy the von Neumann--Mullins relation \cite{1952vonneumann,1956mullins}. The second assumed a finite vertex mobility, and is derived in Section \ref{sec_grain_growth} by considering the forces on a discrete boundary edge. As described in Section \ref{sec_steady_state}, the simulations converged to steady states that do not appear to depend on the initial conditions. Furthermore, simulations with both sets of equations of motion converged to the same steady state, though the convergence is slower in the finite vertex mobility case. This shows that the distance on cloths can be used not only to measure the convergence of a simulations to a steady state, but also to quantify the extent of the differences introduced into simulation results by the use of alternative numerical implementations.

Section \ref{sec_dislocations} describes a model dislocation network where the dislocations are endowed with a constant energy per line length and the network evolves by energy minimization. As with the grain boundary network, simulations beginning from different sets of initial conditions converged toward steady states that were identical within sampling errors. Perhaps more significantly, the distance on cloths shows that the approach from the random graph initial condition to the steady state does not depend on the implementation of a separate topological operation when dislocations intersect, as in Figure \ref{move_3}. That is, the implementation of this specific model system can be significantly simplified and computational requirements can be reduced without measurably changing the statistical local topology of the steady state condition.

A distance on cell complexes is expected to be useful more generally as well. Apart from testing for the convergence of simulations and the invariance of the results to implementation details, the distance allows a meaningful comparison of simulations with experimental observations, the quantification of the variability of cell complexes generated in a particular way, and the iterative modification of a cell complex by continually reducing the distance to an intended target. We sincerely hope that this stimulates further research into statistical topology and its applications to materials science and physics.

\begin{acknowledgments}

The authors would like to thank the Institute for Advanced Study where the original idea occurred. B. Schweinhart was supported by a National Science Foundation Graduate Research Fellowship under Grant No. DGE-1148900, and the Center of Mathematical Sciences and Applications at Harvard University.

\end{acknowledgments}

\bibliography{refs}

\end{document}